\documentclass[sigconf]{acmart}

\usepackage{booktabs} 
\usepackage[utf8]{inputenc}
\usepackage{mathtools}
\usepackage{url}
\usepackage{xcolor}
\usepackage{array}
\usepackage{bbm}
\usepackage{multirow}
\usepackage{xspace}
\usepackage{makecell}
\usepackage{amsfonts}
\usepackage{bm}
\usepackage[caption = false]{subfig}
\usepackage{enumitem}

\usepackage[ruled]{algorithm2e} 

\SetAlFnt{\small}
\SetAlCapFnt{\small}
\SetAlCapNameFnt{\small}
\SetAlCapHSkip{0pt}
\IncMargin{-\parindent}

\acmDOI{}

\acmISBN{}

\acmConference[]{}{}{}
\acmYear{}
\copyrightyear{}

\acmArticle{}
\acmPrice{}

\newcolumntype{L}[1]{>{\raggedright\let\newline\\\arraybackslash\hspace{0pt}}m{#1}}
\newcolumntype{C}[1]{>{\centering\let\newline\\\arraybackslash\hspace{0pt}}m{#1}}
\newcolumntype{R}[1]{>{\raggedleft\let\newline\\\arraybackslash\hspace{0pt}}m{#1}}

\DeclarePairedDelimiter{\norm}{\lVert}{\rVert}
\newcommand{\V}[1]{\mathbf{#1}}

\newcommand{\bfx}{\V{x}}

\newcommand{\bfZ}{\bm{\phi}}
\newcommand{\xad}{\tilde{\bfx}}

\title{DARTS: Deceiving Autonomous Cars with Toxic Signs}

\author{Chawin Sitawarin}
\affiliation{%
  \institution{Princeton University}
  \streetaddress{Department of Electrical Engineering, Princeton University}
  \city{Princeton}
  \state{NJ}
  \postcode{08540}
  \country{USA}}
\authornote{Both authors contributed equally to the paper}
\author{Arjun Nitin Bhagoji}
\affiliation{%
  \institution{Princeton University}
  \streetaddress{Department of Electrical Engineering, Princeton University}
  \city{Princeton}
  \state{NJ}
  \postcode{08540}
  \country{USA}
}
\authornotemark[1]

\author{Arsalan Mosenia}
\affiliation{%
  \institution{Princeton University}
  \streetaddress{Department of Electrical Engineering, Princeton University}
  \city{Princeton}
  \state{NJ}
  \postcode{08540}
  \country{USA}
}

\author{Mung Chiang}
\affiliation{%
  \institution{Purdue University}
  \streetaddress{Department of Electrical and Computer Engineering, Purdue University}
  \city{Princeton}
  \state{NJ}
  \postcode{08540}
  \country{USA}
}

\author{Prateek Mittal} 
\affiliation{%
  \institution{Princeton University}
  \streetaddress{Department of Electrical Engineering, Princeton University}
  \city{Princeton}
  \state{NJ}
  \postcode{08540}
  \country{USA}
}

\begin{abstract}
Sign recognition is an integral part of autonomous cars. Any misclassification of traffic signs can potentially lead to a multitude of disastrous consequences, ranging from a life-threatening accident to even a large-scale interruption of transportation services relying on autonomous cars. In this paper, we propose and examine security attacks against sign recognition systems for \textbf{D}eceiving \textbf{A}utonomous ca\textbf{R}s with \textbf{T}oxic \textbf{S}igns (we call the proposed attacks \textbf{\textit{DARTS}}). In particular, we introduce two novel methods to create these toxic signs. First, we propose \textbf{\signembedding attacks}, which expand the scope of adversarial examples by enabling the adversary to generate these starting from an arbitrary point in the image space compared to prior attacks which are restricted to existing training/test data (In-Distribution). Second, we present the \textbf{\lentprinting attack}, which relies on an optical phenomenon to deceive the traffic sign recognition system. We extensively evaluate the effectiveness of the proposed attacks in both \textit{virtual and real-world settings} and consider both \textit{white-box and black-box threat models}. Our results demonstrate that the proposed attacks are successful under both settings and threat models. We further show that \signembedding attacks can outperform \advtraffic attacks on classifiers defended using the adversarial training defense, exposing a new attack vector for these defenses.
\end{abstract}

\keywords{Adversarial examples, Autonomous cars, Lenticular printing, Security, Sign recognition, Toxic signs, Traffic signs}

\begin{document}

\newcommand{\signembedding}{Out-of-Distribution\xspace}
\newcommand{\lentprinting}{Lenticular Printing\xspace}
\newcommand{\adsign}{Logo\xspace}
\newcommand{\customsign}{Custom Sign\xspace}
\newcommand{\advtraffic}{In-Distribution\xspace}
\newcommand{\sucrate}{\text{VAS\xspace}}
\newcommand{\sucratephy}{\text{SPAS\xspace}}
\newcommand{\detrate}{\text{DR\xspace}}
\newcommand{\auxdata}{auxiliary traffic data\xspace}
\newcommand{\googleimg}{virtual attack with the auxiliary dataset\xspace}
\newcommand{\testimg}{virtual attack with the GTSRB test images\xspace}
\newcommand{\mltscl}{Multi-scale CNN\xspace}
\newcommand{\cnn}{Standard CNN\xspace}

\maketitle

\section{Introduction}\label{sec: intro}
The rapid technological and scientific advancements in artificial intelligence (AI) and machine learning (ML) have led to their deployment in ubiquitous, pervasive systems and applications, such as authentication systems \cite{CABA,AUTH_BOOK}, health-care applications \cite{CARE_1,CARE_2}, and several vehicular services \cite{NVIDIA,bojarski2016end,PROCMOTIVE,PINME}. The ubiquity of ML provides adversaries with both opportunities and incentives to develop strategic approaches to fool learning systems and achieve their malicious goals \cite{6868201,IoT_SURVEY}. A number of powerful attacks on the test phase of ML systems used for classification have been developed over the past few years, including attacks on Support Vector Machines \cite{biggio2014security,moosavi2015deepfool} and deep neural networks \cite{szegedy2013intriguing,goodfellow2014explaining,Carlini16,moosavi2015deepfool,moosavi2016universal,papernot2016limitations}. These attacks have also been shown to work in black-box settings \cite{papernot2016practical,papernot2016transferability,liu2016delving,brendel2017decision,bhagoji2017exploring,chen2017zoo}. These attacks work by adding carefully-crafted perturbations to benign examples to generate adversarial examples. In the case of image data, these perturbations are typically imperceptible to humans. \textit{While these attacks are interesting from a theoretical perspective and expose gaps in our understanding of the working of neural networks, their practical importance has remained unclear in real-world application domains}. 

\begin{figure}[t]
	\centering
	\subfloat[Illustration of Out-of-Distribution evasion attacks on a traffic sign recognition system trained with traffic sign images. Out-of-Distribution attacks enable the adversary to start from anywhere in the space of images and do not restrict her to the training/test data.]{\includegraphics[width=0.49\textwidth]{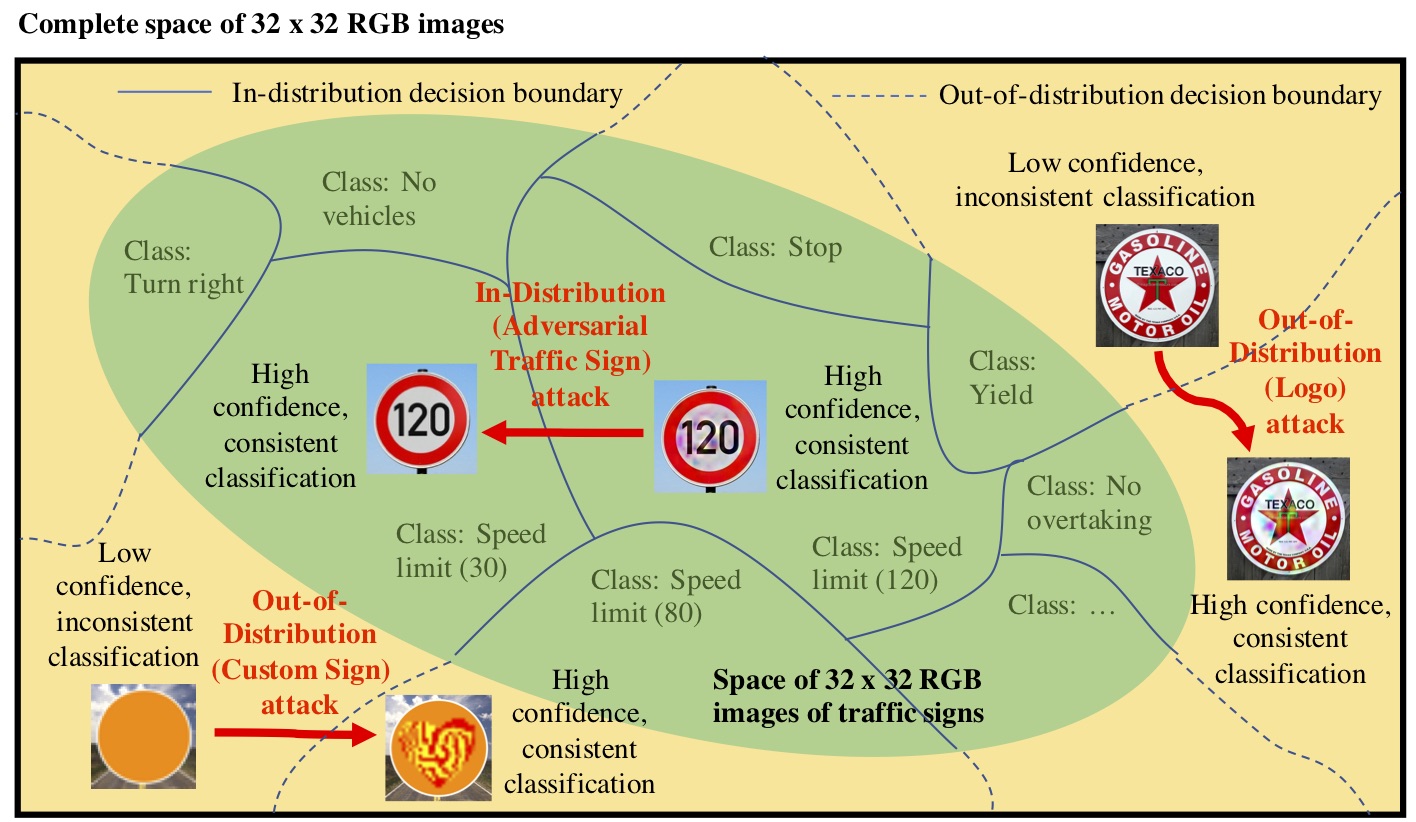}\label{fig: ood_diagram}}
	\newline
	\subfloat[Principle behind an attack based on lenticular printing.]{\includegraphics[width=0.49\textwidth]{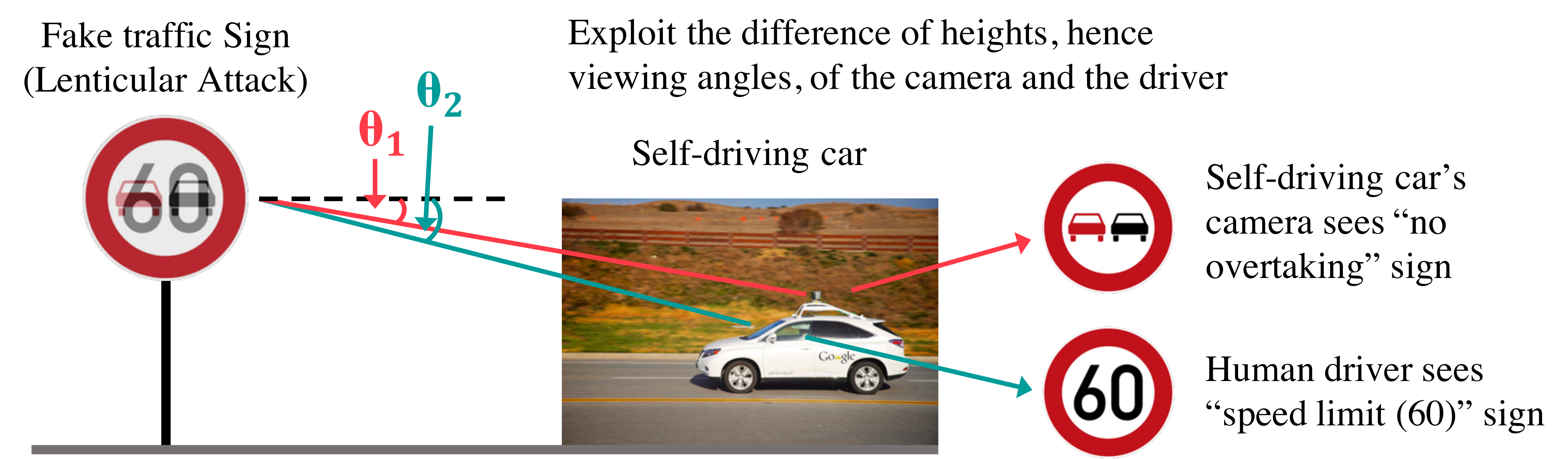}\label{fig: len_attack}}
	\caption{New attack vectors}
	\label{fig: attack_vectors}
	\vspace{-10pt}
\end{figure}


\begin{figure}
	\centering
	\includegraphics[width=0.49\textwidth]{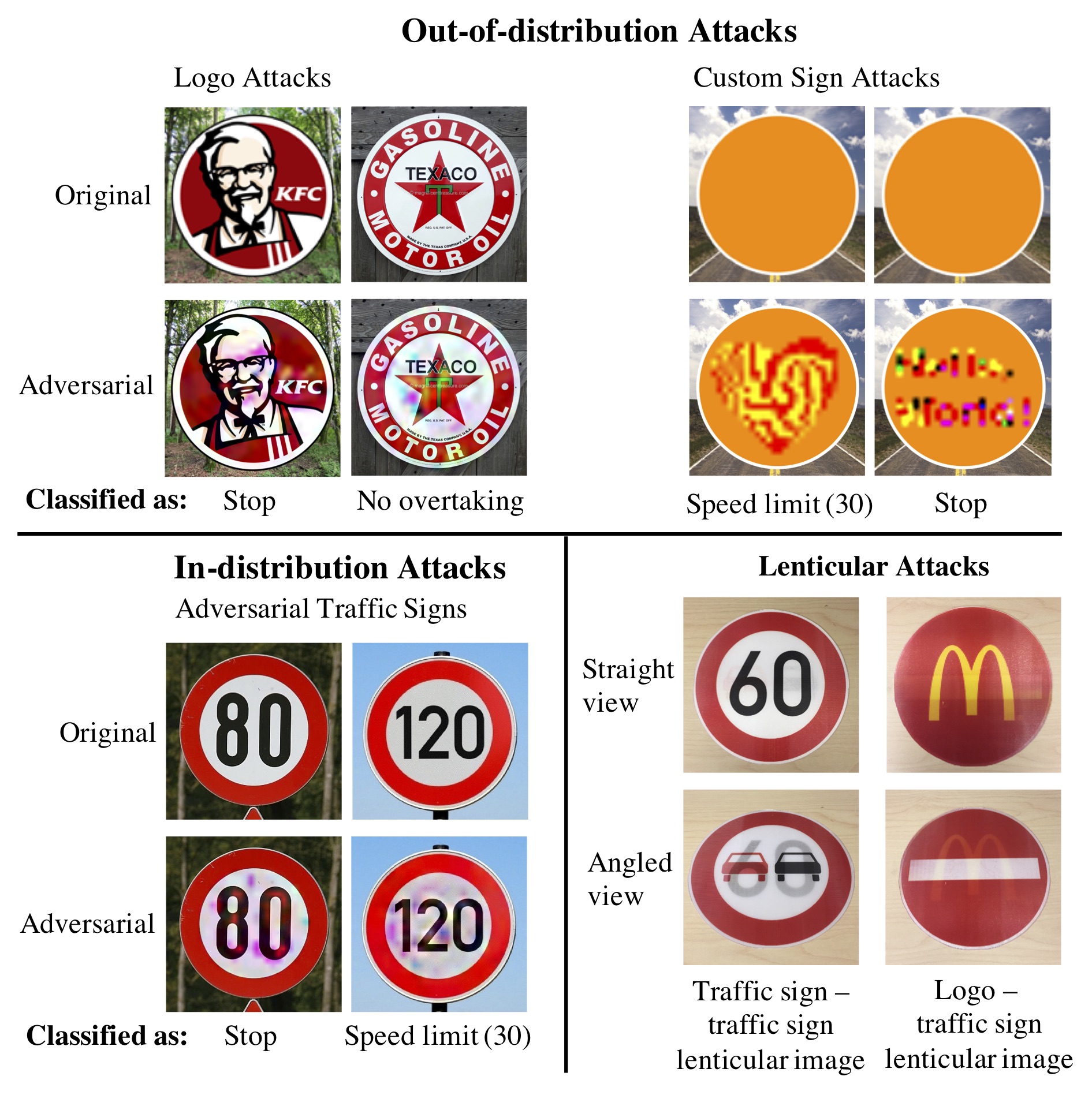}
	\caption{Toxic signs for our traffic sign recognition pipeline generated using the \advtraffic attack, the \signembedding attack (\adsign and \customsign), and the \lentprinting attack. The adversarial examples are \emph{classified as the desired target traffic sign with high confidence} under a variety of physical conditions when printed out. The \lentprinting attack samples flip the displayed sign depending on the viewing angle, simulating the view of the human driver and the camera in the autonomous car.}
	\label{fig: adv_examples}
	\vspace{-15pt}
\end{figure}

A few recent research studies have attempted to provide realistic attacks on ML classifiers, which interact with the physical world \cite{kurakin2016adversarial, sharif2016accessorize}. Kurakin et al. \cite{kurakin2016adversarial} print out virtual adversarial examples, take their pictures, and then pass them through the original classifier. Sharif et al. \cite{sharif2016accessorize} attack face recognition systems by allowing subjects to wear glasses with embedded adversarial perturbations that can deceive the system. However, both these attacks were carried out in \textit{controlled laboratory setting} where variations in brightness, viewing angles, distances, image re-sizing etc. were not taken into account, imposing limitations on their effectiveness in real-world scenarios. Further, they only focus on \textit{creating adversarial examples from training/testing data (e.g., image of different faces in the face recognition system)} to attack the underlying systems. We refer to these as \advtraffic attacks. Athalye et al. \cite{Athalye17} introduced the Expectation over Transformations (EOT) method to generate physically robust adversarial examples and \textit{concurrent} work by Evtimov et al. \cite{Evtimov17} used this method to attack traffic sign recognition systems. A detailed comparison of our work with these is at the end of the introduction.


In this paper, we focus on physically-realizable attacks against sign recognition system utilized in autonomous cars, one of the most important upcoming application of ML \cite{NVIDIA,bojarski2016end, teslaautopilot,applevoxelnet}. We introduce two novel types of attacks, namely, \textit{\signembedding and \lentprinting attacks} that deceive the sign recognition system, leading to potentially life-threatening consequences. Our novel attacks shed lights on \textit{how domain-specific characteristics of an application domain, in particular, autonomous cars, can be exploited by an attacker to target the underlying ML classifier.}    

Our key contributions can be summarized as follows: 

\textit{\textbf{1. Introducing new attack vectors:}} We introduce new methods to create toxic signs that look benign to human observers, at the same time, deceive the sign recognition mechanism. Toxic signs cause misclassifications, potentially leading to serious real-world consequences including road accidents and widespread traffic confusion. We significantly extend the scope of attacks on real-world ML systems in two ways (Figure \ref{fig: attack_vectors}). First, we propose \textbf{\signembedding attack}, which enables the adversary to start from an arbitrary point in the image space to generate adversarial examples, as opposed to prior attacks that are restricted to samples drawn from the training/testing distribution. The proposed attack is motivated by the key insight that autonomous cars are moving through \textit{a complex environment consisting of many objects, which can potentially used by an attacker to create adversarial examples.} We previously provided a high-level description of \signembedding attack in an Extended Abstract \cite{Sitawarin17}. In this paper, we provide an in-depth explanation of this attack and thoroughly examine its effectiveness in various experimental scenarios. Second, we present \textbf{\lentprinting attack}, which relies on an optical phenomenon to fool the sign recognition system. Figure \ref{fig: adv_examples} demonstrates a few toxic signs created by the two above-mentioned attacks.

\textit{\textbf{2. Extensive experimental analysis:}} We evaluate the proposed attacks in both \textit{virtual and real-world settings} over various sets of parameters. We consider both the \textit{white-box threat model (i.e., the adversary has access to the details of the traffic sign recognition system) and the black-box one (i.e., such access is not present)}. We demonstrate that adversarial examples (created from either arbitrary points in the image space or traffic signs) can deceive the traffic sign recognition system with high confidence. Further, we show the attacker can achieve significant attack success rates even in black-box settings. To provide a thorough analysis of the proposed attacks, we also conduct real-world drive-by tests, where a vehicle-mounted camera continuously captures image from the surroundings and offers its data to the sign recognition system (Figure \ref{fig: real-world_result}). We achieve attack success rates in excess of 90\% in the real-world setting with both \signembedding and \advtraffic attacks.
    
\textit{\textbf{3. Studying the effect of \signembedding attacks on state-of-the-art defenses:}} We discuss the limitations of adversarial training based defenses\cite{goodfellow2014explaining} in mitigating the proposed attacks. We show \signembedding attacks, in which the initial image does not come from the underlying training/testing distribution, outperform \advtraffic attacks on adversarial training \cite{goodfellow2014explaining} based defenses in which adversarial examples are created based on the initial training dataset and are considered in the training phase. Known ML-based defenses are intrinsically prone to the \lentprinting attack which has been developed with respect to the physical characteristics of the application domain.

\noindent \textbf{Comparison with Athalye et al. \cite{Athalye17} and Evtimov et al. \cite{Evtimov17}:}
Athalye et al. \cite{Athalye17} introduced the Expectation over Transfomations (EOT) method to generate physically robust adversarial examples. They used the method to generate 3D printed adversarial examples which remain adversarial under a range of conditions. These samples are evaluated on classifiers trained on the Imagenet dataset. In \emph{concurrent} work, Evtimov et al. \cite{Evtimov17} used the EOT method to generate physically robust adversarial `Stop' and `Right Turn' signs in a white-box, \advtraffic setting. They manually crop out the portion of the video frame corresponding to the adversarial example while we use an automated detection pipeline. We further expand the space of adversarial examples in both virtual and real-world settings by introducing \signembedding attacks. We also evaluate the effectiveness of transferability-based black-box attacks for both \advtraffic and \signembedding settings, as well as defenses using adversarial training. We also introduce the completely new attack vector of \lentprinting attacks based on optical phenomena which exist outside the ambit of adversarial examples. 

Our proof-of-concept attacks shed light on fundamental security challenges associated with the use of sign recognition techniques in autonomous cars, paving the way for further investigation of overlooked security challenges in this domain.

\section{Background}\label{sec:background}
In this section, we present relevant background on machine learning systems, the traffic sign recognition pipeline and the adversarial examples and threat models we consider.

\subsection{Supervised machine learning systems }
Machine learning systems typically have two phases, a training phase and a test phase \cite{murphy2012machine}. A classifier $f(\cdot;\theta): \mathcal{X} \rightarrow \mathcal{Y}$ is a function mapping from the domain $\mathcal{X}$ to the set of classification outputs $\mathcal{Y}$. The number of possible classification outputs is then $|\mathcal{Y}|$. $\theta$ is the set of parameters associated with a classifier. $\ell_{f}(\bfx,y)$ is used to represent the loss function for the classifier $f$ with respect to inputs $\bfx \in \mathcal{X}$ and labels $y \in \mathcal{Y}$. The classifier is trained by minimizing the loss function over $n$ samples $\{(\bfx_1,y_1), \ldots (\bfx_n,y_n)\}$ drawn from a distribution $\mathcal{D}$ over the domain $\mathcal{X}$. In the particular case of \emph{traffic sign recognition}, $\mathcal{D}$ is a distribution over images of traffic signs.

Since \emph{deep neural networks} (DNNs) \cite{Goodfellow-et-al-2016} achieve very high classification accuracy in a variety of image classification settings \cite{krizhevsky2012imagenet,taigman2014deepface}, they are being used for the computer vision systems of autonomous cars \cite{applevoxelnet,teslaautopilot}. We thus focus on attacks on DNNs in this paper and define some notation specifically for neural networks. The outputs of the penultimate layer of a neural network $f$, representing the output of the network computed sequentially over all preceding layers, are known as the logits. We represent the logits as a vector $\bfZ^f(\bfx) \in \mathbb{R}^{|\mathcal{Y}|}$. The final layer of a neural network $f$ used for classification is usually a softmax layer. The loss functions we use are the standard \emph{cross-entropy loss} \cite{Goodfellow-et-al-2016} and the \emph{logit loss} \cite{Carlini16}.

\begin{figure}[t]
	\centering
	\includegraphics[width=0.49\textwidth]{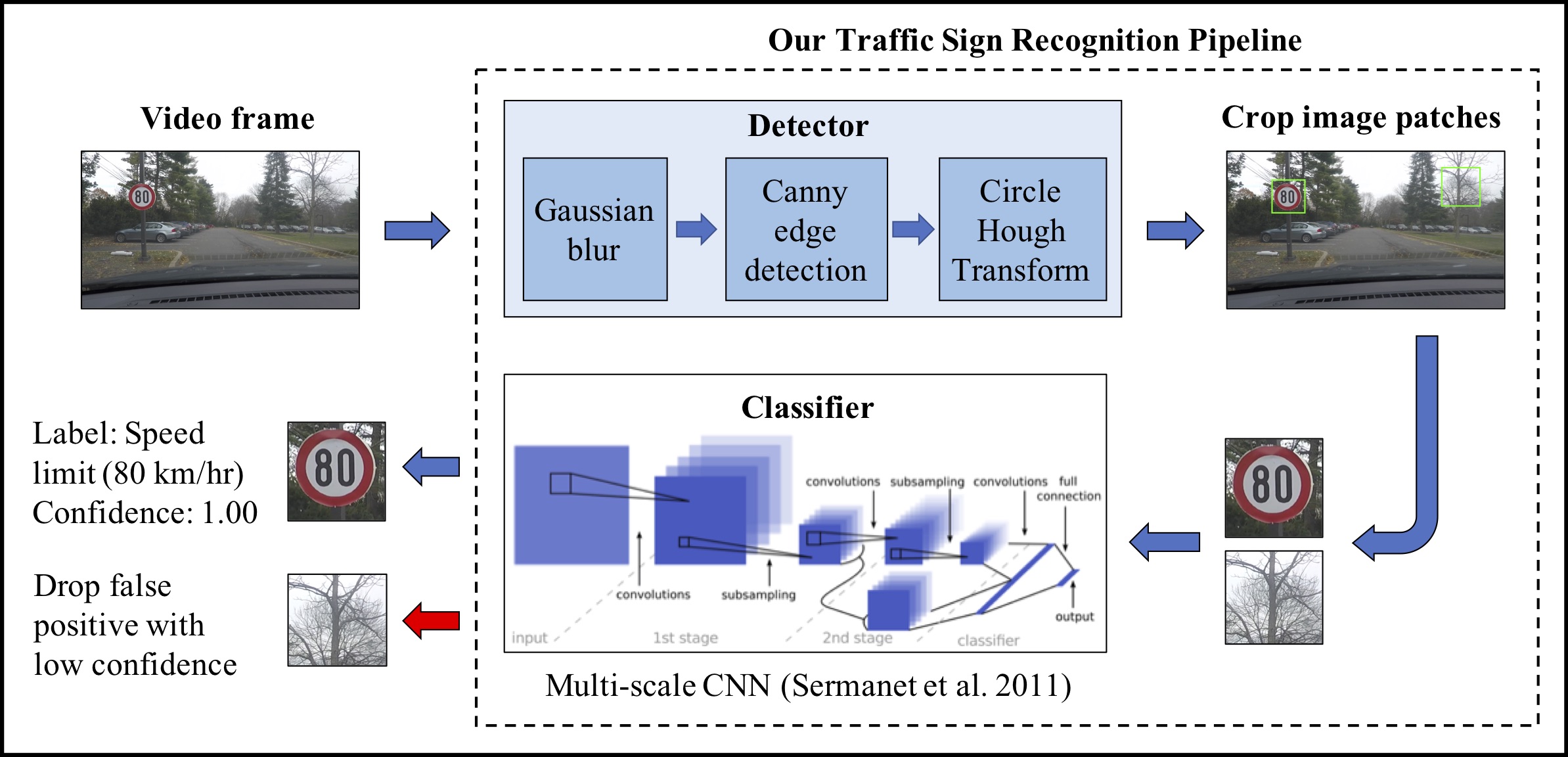}
	\caption{\textbf{Sign recognition pipeline for real-world evaluation}. The pipeline consists of an initial detection phase followed by a multi-scale CNN as a classifier. In the virtual setting, a video frame is replaced by a still image.}
	\label{fig: detect}
	\vspace{-10pt}
\end{figure}

\subsection{Traffic Sign recognition pipeline}
Our traffic sign recognition pipeline (Figure \ref{fig: detect}) consists of two stages: detection and classification. We utilize a commonly used recognition pipeline based on the Hough transform \cite{barrile2012automatic, Yakimov2015Hough, Garrido2005Hough}. The shape-based detector uses the circle Hough transform \cite{HoughCir83:online} to identify the regions of a video frame or still image that contain a circular traffic sign. Before using Hough transform, we smooth a video frame with a Gaussian filter and then extract only the edges with Canny edge detection \cite{canny}. Triangular signs can be detected by a similar method \cite{Yakimov2015Hough}. The detected image patch is cropped and re-sized to the input size of the classifier before it is passed on to the neural network classifier trained on a traffic sign dataset. The classifier outputs confidence scores for all output classes to determine whether the input is a traffic sign and assign its label. The label with the highest confidence is chosen as the final output only if its confidence is above a certain threshold. Images classified with a low confidence score are discarded as false positives for detection. 

\begin{figure*}[t]
	\centering
	\includegraphics[width=0.9\textwidth]{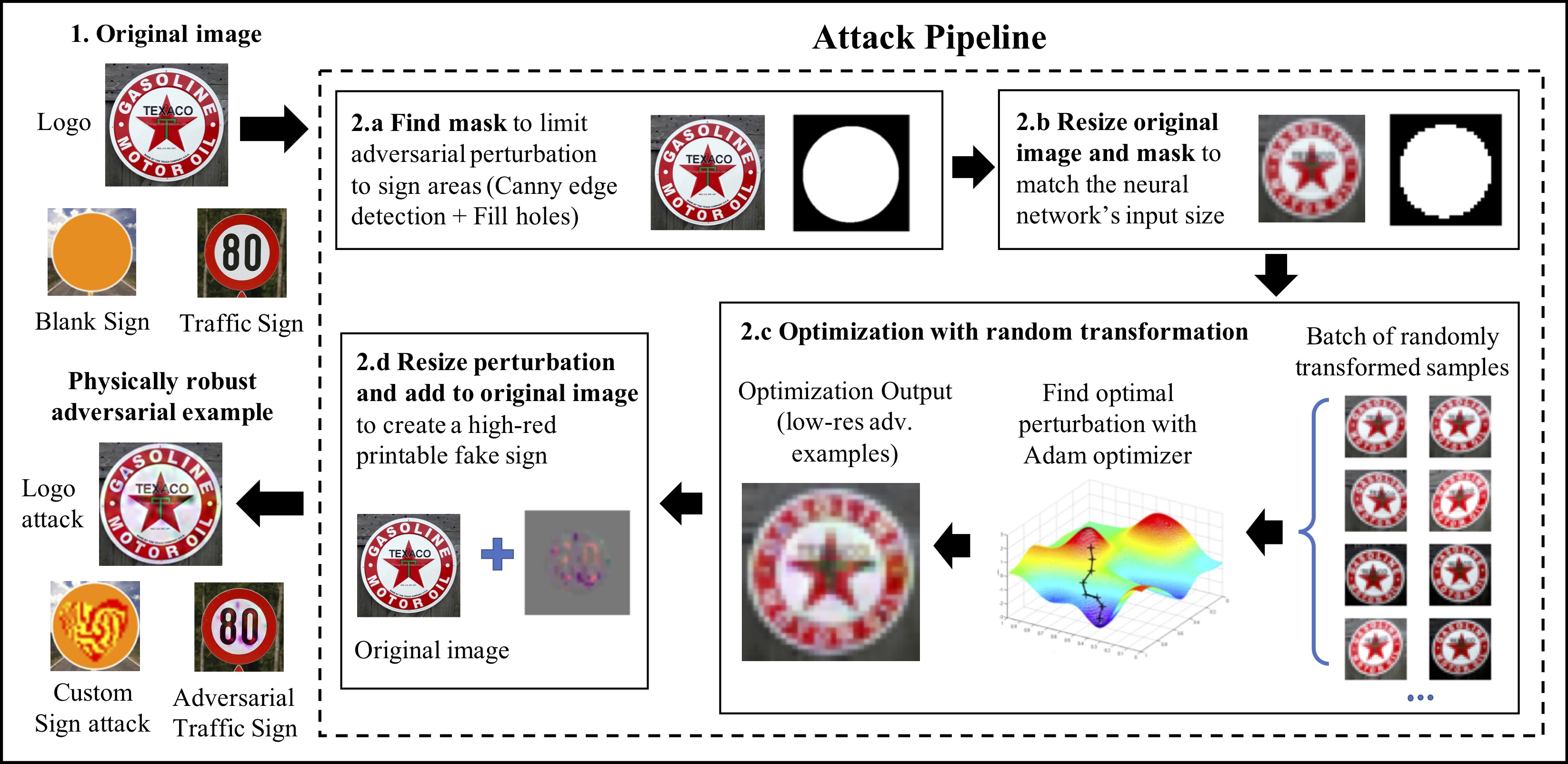}
	\caption{\textbf{Overview of the Attack pipeline}. This diagram provides an overview of the process by which adversarial examples are generated for both the \signembedding attack and \advtraffic attacks.}
	\label{fig: attack}
	\vspace{-10pt}
\end{figure*}

\subsection{Adversarial examples and threat models}
 Our focus is on attacks during the test phase, which are typically known as \emph{evasion attacks}. These have been demonstrated in the virtual setting for a number of classifiers \cite{biggio2013evasion,szegedy2014intriguing,goodfellow2014explaining,Carlini16,moosavi2015deepfool,papernot2016limitations}. These attacks aim to modify benign examples $\bfx \in \mathcal{X}$ by adding a perturbation to them such that the modified examples $\xad$ are \emph{adversarial}, i.e. they are misclassified by the ML system in a targeted class (targeted attack), or any class other than the ground truth class (untargeted attack). In the case of attacks on the computer vision systems of autonomous cars,we focus entirely on \emph{targeted attacks} since these are more realistic from an attacker's perspective. To generate a \emph{targeted} adversarial sample $\xad$ starting from a benign sample $\bfx$ for a classifier $f$, the following optimization problem \cite{Carlini16} leads to state-of-the-art attack success rates in the virtual setting:
\begin{align}
	\min \quad & d(\xad, \bfx) + \lambda \ell_f(\xad,T), \label{eq: cw_opti} \\
	\text{s.t.} \quad & \xad \in \mathcal{H} \nonumber.
\end{align}
We use this attack as a baseline. Here $d$ is an appropriate distance metric for inputs from the input domain $\mathcal{X}$ (usually an $L_p$ norm), $T$ is the target class and $\mathcal{H}$ is the constraint on the input space. $\lambda$ controls the trade-off between minimizing the distance to the adversarial example and minimizing the loss with respect to the target. In essence, the optimization problem above tries to find the closest $\xad$ to $\bfx$ such that the loss of the classifier at $\xad$ with respect to the target $T$ is minimized. For a neural network, $\ell_f(\cdot,\cdot)$ is typically highly non-convex, so heuristic optimizers based on stochastic gradient descent have to be used to find local minima \cite{Carlini16}. 

For traffic sign recognition systems, the method described above produces adversarial examples which do not work well under conditions encountered in the real world such as variation in brightness, viewing angles, image re-sizing etc.. That is, if the examples generated using this method were directly used in Step 3 (mount sign and drive-by test) of Figure \ref{fig: attack}, they would not perform well. In light of this, we incorporate these conditions directly into a modified optimization problem used in Section \ref{sec: adv_examples} to generate physically robust adversarial examples. Similar approaches to this problem have been taken in Athalye et al. \cite{Athalye17} as well as in concurrent work by Evtimov et al. \cite{Evtimov17}.

\subsubsection{Threat models}  \label{subsubsec: threat_models}
We consider two commonly used threat models for the generation of adversarial examples against deep neural networks: \emph{white-box} and \emph{black-box}. The black-box setting also applies to the \lentprinting attack.

\noindent \textbf{White-box:} In the white-box setting, we assume that the adversary has complete access to the target model $f$ including its architecture and weights. We briefly justify the powerful attacker considered in the white-box setting. Consider an attacker who wishes to cause an autonomous car to detect and misclassify a sign in its environment. It is conceivable that the attacker can purchase or rent a vehicle of the kind that is to be attacked, and `reverse engineer' the classifier by querying it on an appropriate dataset to train a surrogate classifier that closely mimics the target classifier~\cite{tramer2016stealing}. Further, direct query-based attacks can be as powerful as white-box attacks \cite{bhagoji2017exploring,chen2017zoo}. Thus, the white-box setting is an important one to consider.

\begin{figure*}[t]
	\centering
	\subfloat[Classification of \adsign attack examples. The adversarial examples are classified with high confidence as a real traffic sign, in spite of being out of the dataset.]{\includegraphics[width=0.8\textwidth]{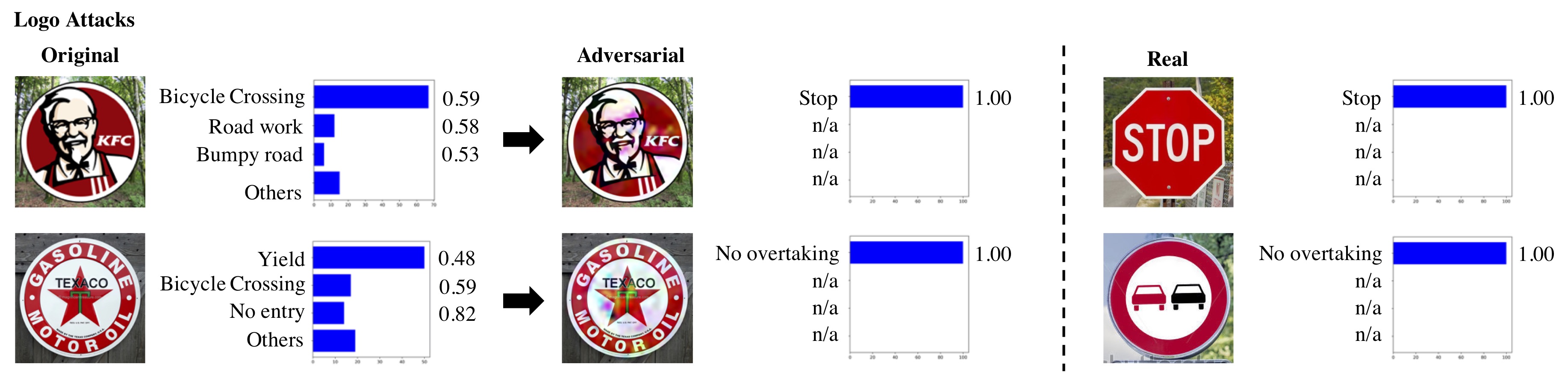}\label{fig: logo}}
	\newline
	\subfloat[Classification of \customsign attack examples. The adversarial examples are classified with high confidence as a real traffic sign, in spite of being custom made signs.]{\includegraphics[width=0.8\textwidth]{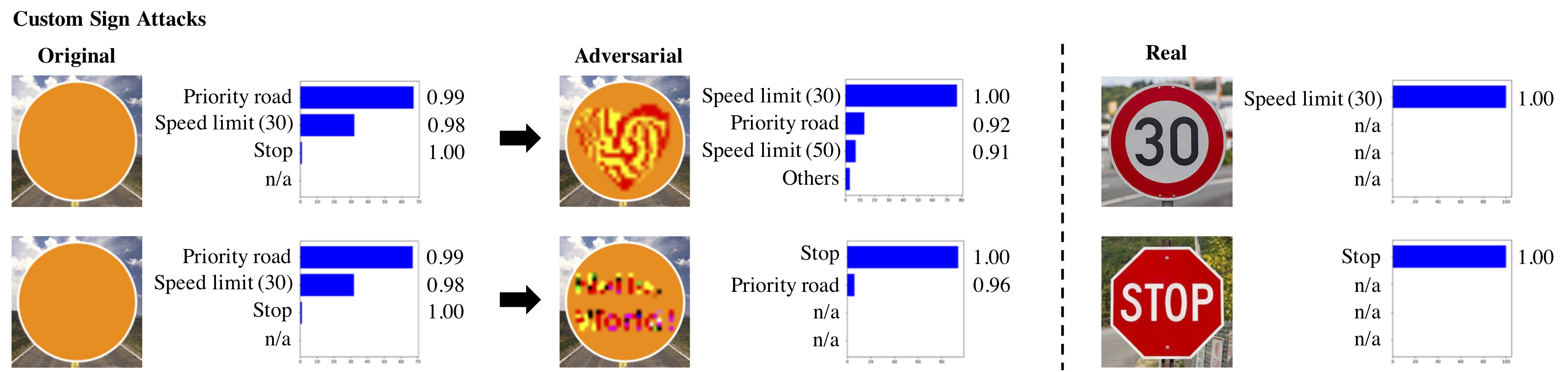}\label{fig: custom}}
	\newline
	\caption{Frequency of the top-3 labels the given images are classified as under 100 different randomized transformations. The numbers beside each bar on the bar charts provides the average confidence of the top classification outcomes over the 100 randomized transformations.}
	\label{fig: virtual_adv_samples}
	\vspace{-10pt}
\end{figure*}

\noindent \textbf{Black-box (no query access):} In this setting, the adversary does not have direct access to the target model. We do not even assume query access as in previous work \cite{brendel2017decision,chen2017zoo,bhagoji2017exploring}. In this setting then, black-box attacks rely on the phenomenon of transferability \cite{szegedy2013intriguing,papernot2016transferability,papernot2016practical,liu2016delving}, where adversarial examples generated for a model trained locally by the attacker, remain adversarial for a different, target model. The \lentprinting attack naturally operates in the black-box setting since it relies on an optical phenomenon and not on the internal structure of the classifier.

\section{Attacks: Adversarial Examples} \label{sec: adv_examples}
In this section, we present new methods to generate physically robust adversarial examples. The main aim for these examples is to be \emph{classified consistently as the desired target class under a variety of real-world conditions} such as variations in brightness, viewing angle, distance and image re-sizing. In particular, we introduce \emph{\signembedding attacks}, which modify arbitrary, innocuous signs such that they are detected and classified as potentially dangerous traffic signs in the attacker's desired class. This attack \textit{greatly enlarges the possible space of adversarial examples} since the adversary now has the \textit{ability to start from any point in the space of images to generate an adversarial example}. We also examine the \advtraffic attack which generates adversarial examples starting from existing traffic signs. We then describe the pipeline we use to ensure these generated adversarial examples are classified consistently even under randomized input transformations which we use to model real-world conditions that may be encountered.

\subsection{Attack overview} \label{subsec: attack_overview}
We first provide an overview of our attack pipeline from Figure \ref{fig: attack} and then describe each of its components in detail in the following subsections. Our pipeline has three steps:\\

\noindent  \textbf{Step 1.} Obtain the original image $\bfx$ and choose target class $T$ that the adversarial example $\xad$ should be classified as. In our attacks, $\bfx$ can either be an \textit{in-distribution image} of a traffic sign or an \textit{out-of-distribution} image of an ad sign, logo, graffiti etc.  \\
\noindent \textbf{Step 2.} Generate the digital version of the physically robust adversarial example as follows:

\indent \textbf{1.} Generate mask $M$ for the original image (A mask is needed to ensure that the adversary's perturbation budget is not utilized in adding perturbations to the background.)

\indent \textbf{2.} Re-size both the original image and the mask to the input size of the target classifier\footnote{This is $32\times32$ pixels for the classifiers we use, described in Table \ref{tab: eval_summary}}.

\indent \textbf{3.} Run the physically robust adversarial example optimization problem from Equation \ref{eq: opt_problem} to obtain the perturbation $\bm{\delta}$. The optimization problem includes randomized brightness, perspective (includes rotations and shearing) and size variations while generating the adversarial example to ensure it remains adversarial under real-world conditions.

\indent \textbf{4.} Re-size the output perturbation $\bm{\delta}$ and add it to the original image.

\noindent \textbf{Step 3.} Print and test the generated adversarial example $\xad$ for robustness in real-world conditions.

\subsection{Step 1: Choosing the input} \label{subsec: attack_types}
Now, we describe two different attack modes in which an adversary can operate to generate adversarial examples: (1) the \signembedding attack, where the adversary is free to generate adversarial examples from any innocuous elements in the environment such as advertisement signs, drawings, graffiti etc. and (2) the \advtraffic attack, where the adversary modifies existing traffic signs to make them adversarial. The second setting is similar to the one considered in most previous work on generating adversarial examples \cite{goodfellow2014explaining,Carlini16}.

\subsubsection{\signembedding attacks} \label{subsubsec:signembedding}
We propose a novel attack based on the concept of adversarial examples by exploiting the fact that any image in the input domain of the classifier can be made adversarial by adding an imperceptible perturbation generated by an optimization problem such as Equation \ref{eq: opt_problem} which also ensures physical robustness. Since the classifier is only trained on images of traffic signs, it can only be expected to reliably classify other traffic signs, which are effectively, \emph{in distribution} for that classifier. However, the fact that it provides a classification outcome for \emph{any input image}, represents a security risk which we exploit in our attacks. In particular, we start with an \emph{out-of-distribution image} (not a traffic sign) and generate a targeted adversarial example from it. Here, we demonstrate two possible instantiations of the \signembedding attack.

\textbf{ \adsign attacks:}  In this attack, images of commonly found logos are modified such that they are detected and classified with high confidence as traffic signs (Figure \ref{fig: logo}). Since these logos are omnipresent, they allow an adversary to carry out attacks in real-world settings such as city streets. In this scenario, the attack pipeline from Section \ref{subsec: attack_overview} is used and the adversarial perturbation is constrained to be as small as possible while still being effective under transformations.

\textbf{ \customsign attacks:}  In this attack, the adversary creates a custom sign that is adversarial starting from a blank sign (Figure \ref{fig: custom}). Any mask corresponding to graffiti, text etc. on blank signs can lead to the embedding of adversarial traffic signs in inconspicuous, graffiti-like objects in the environment. This allows the adversary to create adversarial examples in any setting by using a mask to create images or text that are appropriate for the surroundings. In this attack, the original sign is a solid color circular sign and the norm of the perturbation is not penalized by the optimization problem but only constrained by the shape of the mask. This allows the adversary to draw almost any desired shape on a blank sign and the optimization will try to fill out the shape with colors that make the sign classified as the target class.

Under ordinary conditions when no adversarial examples are present, the false recognition of objects which are not traffic signs does not affect the traffic sign recognition system since (i) the confidence scores corresponding to the predicted labels of these objects are usually low; (ii) these circular objects are not consistently classified as a certain sign. The predicted label changes randomly as the background and viewing angle varies across multiple frames in the video.

Therefore, a traffic sign recognition system, including ours, can choose to treat any detection with these two properties as an erroneous detection by setting the confidence threshold close to 1 and/or ignoring objects that are inconsistently classified. On the other hand, the generated adversarial examples are \emph{classified consistently as target traffic signs with high confidence under varying physical conditions} which we demonstrate experimentally below.

\textbf{Confirming hypothesis about classifier confidence:} 
To confirm our earlier hypotheses with regard to the confidence of classification for \signembedding images and their adversarial counterparts, we display some adversarial examples, their classification and confidence in Figure \ref{fig: virtual_adv_samples}. Experimental details are in Section \ref{sec: white-box}. 

We apply randomized transformations to create 100 images for each of the Logo signs. The "Original" column of Figure \ref{fig: logo} shows that the logo signs are classified as different classes depending on the transformation and with low confidence on average. As a comparison, the ``Real" column of Figure \ref{fig: virtual_adv_samples} shows that real traffic signs  are consistently classified as the correct label with probability very close to 1. The ``Adversarial" column demonstrates that the generated adversarial examples are \textit{consistently classified with high confidence} in the desired target class. In a sense, the adversarial logo signs and the real traffic signs are equivalent from the perspective of the classifier.

Similarly, the the \customsign attack samples are mostly classified as the target class with high confidence under 100 different randomized transformations. Note that even though the original solid color signs shown in the ``Original" column of Figure \ref{fig: custom} are classified with high confidence, their classified labels are arbitrary and not under the control of the adversary. They are highly dependent on the signs' orientation and brightness and cannot be used as reliable adversarial examples. For example, a slight change in the camera angle may yield different labels which is an undesirable effect from the adversary's perspective.

\noindent \textbf{Note:} \signembedding attacks can be carried out against any classifier in any application domain. We choose to focus on traffic sign recognition systems since they provide a compelling setting where such attacks are plausible and effective. Another application domain of interest is the computer vision systems of Augmented Reality systems. In certain other settings, such as fooling content moderation systems \cite{bhagoji2017exploring}, \signembedding attacks may be of limited interest.

\subsubsection{\advtraffic attack} \label{subsubsec: adv_traffic}
In this attack, images of \emph{traffic signs} are modified using imperceptible perturbations such that they are classified as a different traffic sign. This attack is similar to most attacks carried out in most previous work in both the virtual \cite{szegedy2013intriguing,goodfellow2014explaining,papernot2016limitations, Carlini16, moosavi2015deepfool} and physical \cite{kurakin2016adversarial,sharif2016accessorize, Evtimov17, Athalye17} settings. We include it here to demonstrate that our pipeline works in this setting as well.


\begin{figure}
    \centering
    \includegraphics[width=0.49\textwidth]{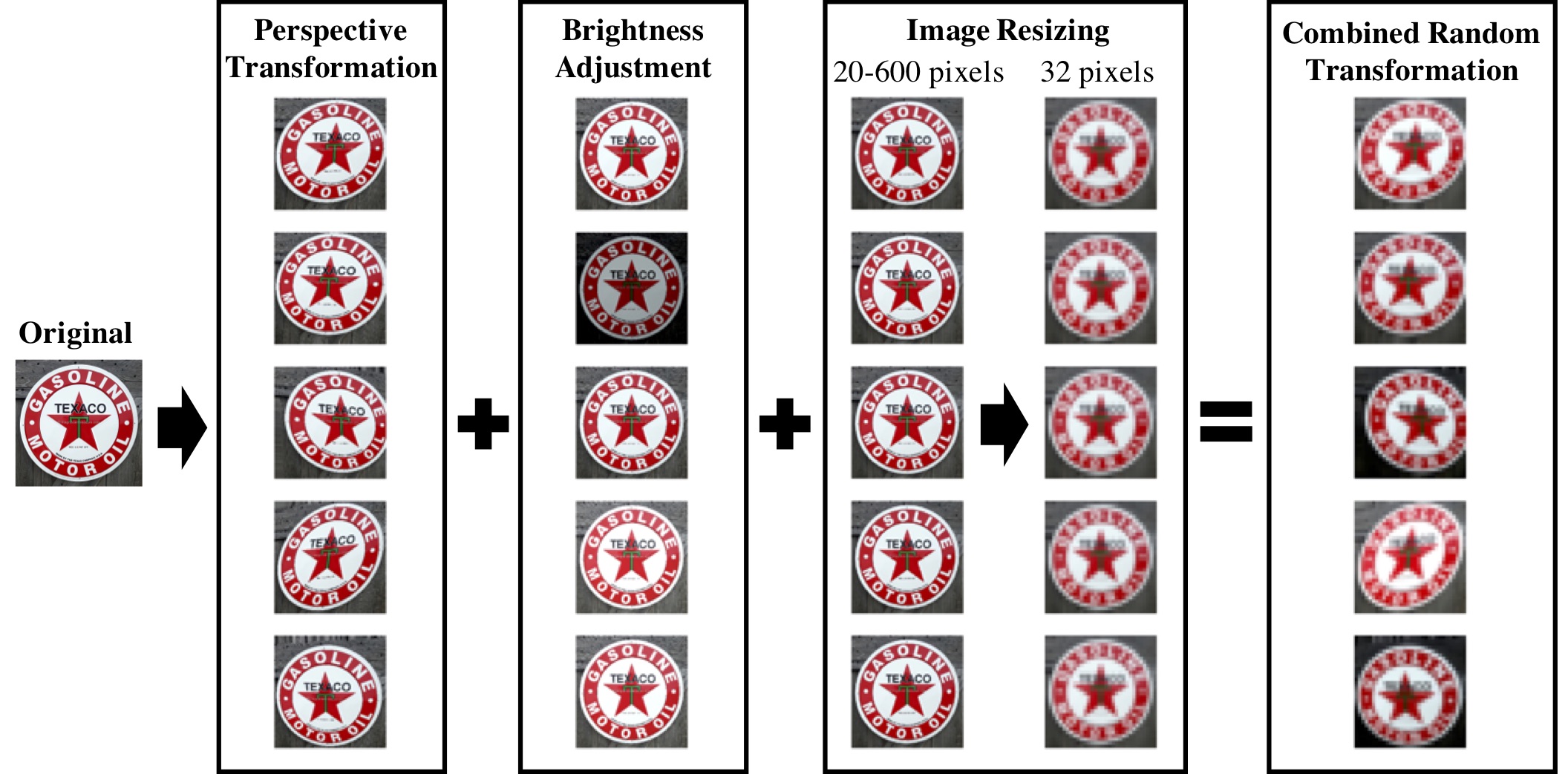}
    \caption{Transformations used during both training and evaluation to simulate real-world conditions. The final column of images displays the type of images that are included in the sum in Eq. \eqref{eq: opt_problem}.}
    \label{fig: transformations}
    \vspace{-15pt}
\end{figure}

\subsection{Step 2: Robust adversarial example generation}
In this section, we describe our methodology for generating robust, physically realizable adversarial examples starting from either of the inputs described in the previous section.

\subsubsection{Optimization problem} \label{subsubsec: opti_problem}
Our adversarial example generation method involves heuristically solving a non-convex optimization problem using the Adam optimizer \cite{kingma2014adam} to find the optimal perturbation $\bm{\delta}$. The problem set-up is adapted from the general concept of expectation over transformations \cite{Athalye17}. An updating gradient is averaged over a batch of randomly transformed versions of the original image \cite{Sharif16,Evtimov17}. The robust adversarial example can be written as a solution to the minimization problem given below for any input $\bfx$: 
\begin{align} \label{eq: opt_problem}
\begin{split}
& \min_{\bm{\delta}\in\mathbb{R}^n} \quad c\cdot\frac{1}{B}\sum_{i=1}^{B} \left[ F(\tau_i(\bfx+M\cdot\bm{\delta})) \right] + \max(\norm{\bm{\delta}}_p, L)
\end{split}
\end{align}
where $F(\bfx)=\max(\max_{j \neq T}\{\bfZ(\bfx)_j\} - \bfZ(\bfx)_t, -K)$ is the logit loss\cite{Carlini16} and $\bfZ(\bfx)_j$ is the $j^{\text{th}}$ logit of the target network. The constant $K$ determines the desired objective value controls the \textit{confidence score} of the adversarial example. $M$ is a mask or a matrix of 0s and 1s with the same width and height as the input image which is multiplied element-wise with the perturbation ($M\cdot\bm{\delta}$) to constrain the feasible region to the sign area.  $\tau_i:\mathbb{R}^n\to\mathbb{R}^n$ is a transformation function mapping within the image space ($n=32\times32\times3$). The transformations $\tau_i$ are chosen to be differentiable so that gradients can be propagated back to the original image. 

The objective value is penalized by a $p$-norm of the perturbation $\bm{\delta}$ in order to keep the adversarial perturbation imperceptible by humans, and the constant $c$ is adjusted accordingly to balance between the loss and penalty terms.  We introduce an additional constant $L$ to explicitly encourage the norm of the perturbation to be at least $L$ since \emph{overly small perturbations can disappear or be rendered ineffective in the process of printing and video capturing}. In practice for the \customsign attack, the same optimization problem is used by setting $c$ and $L$ to some large numbers so that the optimization will focus on minimizing the loss function without penalizing the norm of the perturbation.

\noindent \textbf{Summary:} The optimization problem tries to minimize the average of the objective function with respect to a perturbation $\bm{\delta}$ restricted to an area determined by a mask $M$ evaluated on a batch of $B$ images which are generated from a set of transformation functions $\tau_1,...,\tau_B$. 

\subsubsection{Image selection and re-sizing}
We use a single RGB image, re-sized to the input size of the classifier. We use either bilinear interpolation or nearest neighbours for re-sizing. This input image is then transformed by the set of differentiable and predefined transformations with randomized parameters. The attacker has a low cost with regards to constructing an adversarial example since only a single image is needed to construct robust adversarial examples. In the case of \advtraffic attacks, the adversary only needs one image of the same class as the original sign which can be easily found on the internet. The image also does not need to be a real photograph; as we show in our \signembedding attacks, any logo or schematic drawing can be used as well.

\subsubsection{Mask generation}
A mask is a mapping from each pixel of the image to 0 or 1. Background pixels are mapped to 0 while pixels from the  sign area are mapped to 1. To create a \textit{mask}, we employ a simple image segmentation algorithm \footnote{Code adapted from \cite{ImageSeg43:online}} using Canny edge detection \cite{canny1987computational} to outline the boundary of the sign and binary dilation to fill in the hole. This algorithm works well on an image of a sign of any shape, given that it is of a high resolution. 


\subsubsection{Transformations} \label{subsubsec:transforms}
Here, we describe the set of image transformations used in the optimization which are chosen to virtually generate a batch of input images resembling the varying real-world conditions under which the generated physical adversarial example is expected to be effective. We included three different differentiable transformations in our experiments: (1) perspective transform, (2) brightness adjustment and (3) re-sizing, each of which is randomized for a particular image while generating an adversarial example as well as while evaluating it (see metrics used in Section \ref{subsubsec: eval_metrics}). Examples of transformations used for both the optimization and evaluation phases are shown in Figure \ref{fig: transformations}.

\textbf{Perspective transformation:} Orientations of a sign that appears on an image vary between camera angles, distances and relative heights. These orientations can be captured by perspective transformation, an image transformation that maps each of the four corners of a 2D image to a different point in the 2D space. Any perspective transformation can be characterized by eight degrees of freedom which can be represented by a $3\times3$ \textit{projective transform matrix} with the last entry being set to $1$. Other common image transformations such as rotation, shearing or scaling are subsets of the perspective transformation.

\textbf{Brightness adjustment:}  Naturally, cars drive any time of the day so an adversary might need the fake sign to be able to fool the system under different lighting conditions. The simplest way to simulate settings with different amounts of light or brightness is to add a constant value to every pixel and every channel (R, G, B) of an image and then clip the value to stay in the allowed range of $[0, 1]$. This transformation consisting of an addition and clipping is differentiable almost everywhere.

\textbf{Image re-sizing:}  Due to varying distances between a camera and a sign, an image of the sign detected and cropped out of the video frame could have different sizes (equivalent to number of pixels). Further, after generation, adversarial perturbations are up-sampled to match the original image size. Subsequently, when the adversarial sign is detected on the camera, it is cropped out and down-sampled back to $32\times32$ pixels for the classifier. This re-sampling process could blur or introduce artifacts to the image, diminishing the intended effect of adversarial perturbations. We use a \emph{nearest neighbours} re-sizing technique for the mask and \emph{bilinear interpolation} for all other images. 

The randomized transformation $\tau$ used during both the generation and evaluation of adversarial examples is a composition of the three transformations mentioned above. The degree of randomness of each transformation is controlled by a set of tunable parameters to keep them in a range that provides realistic simulations.\\

\noindent \textbf{Parameter choices:} Our optimization program introduces a number of parameters ($c, K, L, T, \theta, \text{step size}, \text{choices of norm}$) that can be fine-tuned to control the trade-off between robustness of adversarial examples and their visibility to humans. A detailed analysis of the procedure we followed to optimize these parameters in tandem to achieve high attack success rates in practice is in Appendix \ref{app: param_tune}.


\section{Attacks: Lenticular Printing} \label{sec:lent}
In this section, we describe a novel attack on traffic sign recognition systems that we call the \lentprinting attack. 

\begin{figure}[t]
	\centering
	\subfloat[Illustration of the process of generating a lenticular image and its angle-dependent appearance.]{\includegraphics[width=0.49\textwidth]{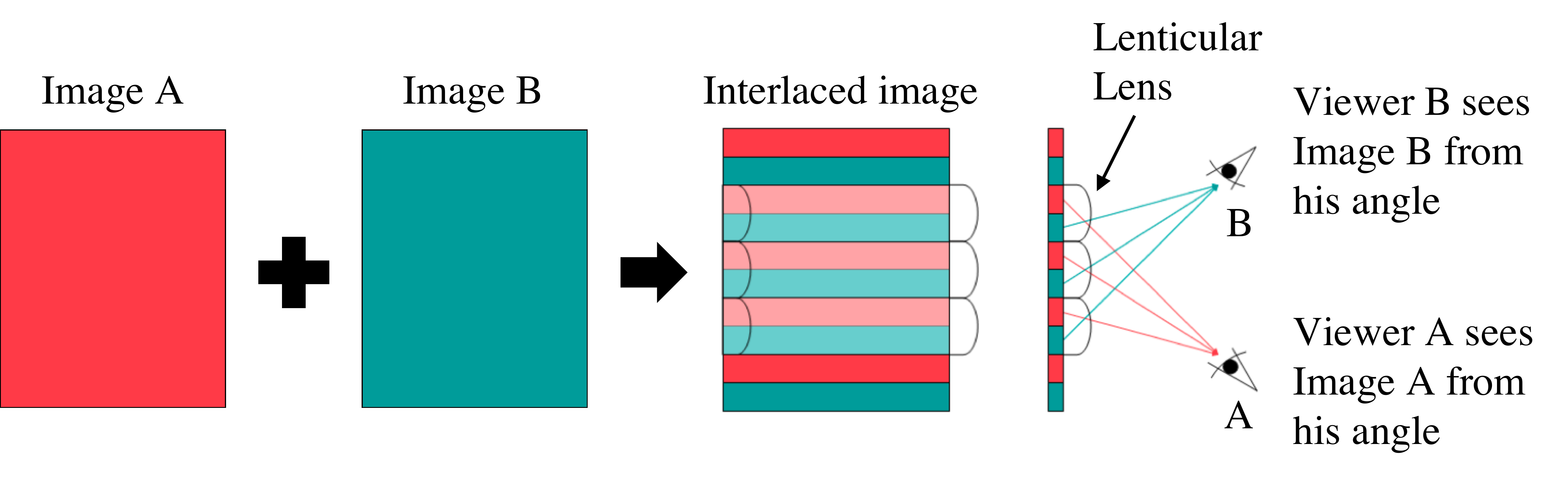} \label{fig: lenticular}}
    \newline
	\subfloat[Several parameters ($p$, $r$, and $h$) determine the full angle of observation ($O$)]{\includegraphics[scale=0.3]{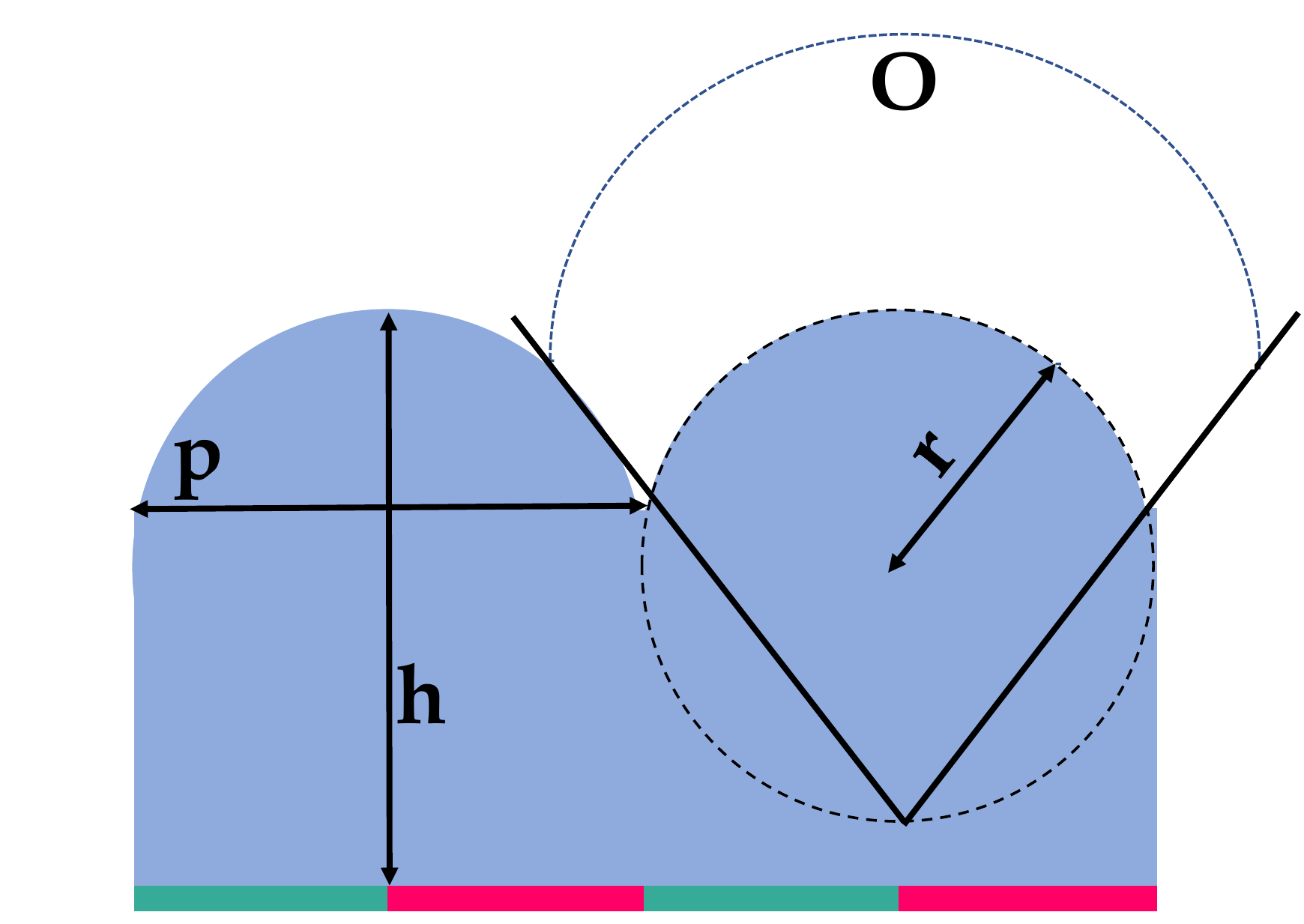} \label{fig: para}}
	\caption{Concepts underlying the working of the \lentprinting attack.}
	\vspace{-15pt}
\end{figure}

\begin{table*}
	\centering
	\resizebox{0.99\textwidth}{!}{
		\begin{tabular}{c|c|c|c}
			\toprule
			Component & Name & Description & Usage \\ \midrule
			\multirow{5}{*}{Datasets} & GTSRB \cite{Stallkamp2012} &  \begin{tabular}{@{}c@{}} 51,839 $32 \times 32 $ RGB images of 43 types of traffic signs (39,209 for training, 12,630 \\ for testing) (Figure \ref{fig: gtsrb}) Training set is augmented to 86,000 samples (see Appendix \ref{app: data_details}). \end{tabular} & Training and validating both classifiers \\
			& GTSDB \cite{Houben2013} & 900 $1360 \times 800$ RGB images of traffic signs in real-world conditions & Evaluating detector+classifier pipeline \\
			& Auxiliary traffic data & 22 high-resolution RGB images of traffic signs (Figure \ref{fig: auxdatasign}) & Generating real-world examples for the \advtraffic attack \\
			& \adsign & 7 high-resolution RGB images of popular logos (Figure \ref{fig: logosign}) & Generating real-world examples for the \signembedding attack\\
			& \customsign & \begin{tabular}{@{}c@{}} 3 blank circular signs (orange, white and blue) \\ combined with 6 custom masks to constrain perturbations (Figure \ref{fig: customsign}) \end{tabular} & Generating real-world examples for the \signembedding attack \\ \midrule
			\multirow{2}{*}{Classifiers} & \mltscl \cite{LeCun2011mltscl} & \begin{tabular}{@{}c@{}} Multi-scale CNN with 3 convolutional layers and 2 fully connected layers \\ \textbf{98.50\%} accuracy on GTSRB, \textbf{77.1\% mAP} on GTSDB, 100\% accuracy on \auxdata \end{tabular}  & Main target network (virutal and real-world) \\
			& \cnn & \begin{tabular}{@{}c@{}} Conventional CNN with 4 convolutional layers and 3 fully connected layers \\ \textbf{98.66\%} accuracy on GTSRB, \textbf{81.5\% mAP} on GTSDB, 100\% accuracy on \auxdata \end{tabular}  & Evaluating transferability-based black-box attacks \\ \bottomrule
			
		\end{tabular}
	}
	\caption{Summary of classifiers and datasets used for experimental evaluation.}
	\label{tab: eval_summary}	
	\vspace{-15pt}
\end{table*}

\textbf{Key idea:} \lentprinting attack is \textit{motivated by the key insight that the human driver and the vehicle-mounted camera observe the environment from two different observation angles} (Figure \ref{fig: attack_vectors}). To deceive the sign recognition system utilized in autonomous cars, we create special traffic signs that appear differently from different observation angles (Figure \ref{fig: adv_examples}). In particular, we exploit a multi-step process, known as lenticular printing, that involves creating a special image from at least two existing images, and combining it with an array of magnifying lenses (Figure \ref{fig: lenticular}). Lenticular printing relies on an optical phenomenon and has been traditionally used in photography and visual arts to generate 2-D images that can offer an illusion of depth and be changed as the image is viewed from different angles.

\textbf{Choosing an appropriate lens array:} Here, we briefly describe how an attacker can use this technique to create malicious traffic signs. To maximize the possibility of observing two different images from two different angles, the attacker should interlace images so that the width of each row of an image (red and green rows in Figure \ref{fig: lenticular}) is equal to half of the width of each lens in the lens array, i.e., $Width_{rows}=p/2$ , where $p$ is shown in \ref{fig: para}. For a successful attack, the attacker must ensure that the difference between the observation angle of the driver and the observation angle of the camera always remains between $O/4$ and $O/2$ (i.e., $O/4 \leq |\theta_1-\theta_2| \leq O/2$), where $\theta_1-\theta_2$ are shown in Figure \ref{fig: len_attack} and $O$ is the full angle of observation of a lenticular lens (Figure \ref{fig: para}) and can be determined as follows:
\begin{align}
    O=2(sin^{-1}(\frac{p}{2r})-
    sin^{-1}(\frac{n*sin(sin^{-1}(\frac{p}{2r})-tan^{-1}(\frac{p}{h}))}{n_a})),
\end{align}

where $n_a$ is 1.003 (the index of refraction of air), $n$ is the index of refraction of the lens and all other parameters are shown in Figure \ref{fig: para}. In our experiments, we use commonly-available lens with a $49\sim50$-degree full angle of observation (i.e., $O\approx 50 \Rightarrow 12.5<|\theta_1-\theta_2|<25$ should be maintained during driving). 

\textbf{Creating malicious signs:} Industrial quality lenticular printing involves specialized machinery and trained personnel to operate it. Nevertheless, a simple lenticular image can also be produced without the need for specialized equipment by using either an inkjet or a laser color printer and a lenticular lens. We have created lenticular images using a two-step procedure: (i) we obtain two images of the same dimensions, one of which is a standard traffic sign (the adversary's desired target) and the other one can be a logo, a custom sign, or another traffic sign. These just have to be of the same dimensions so that they can be interlaced with the desired traffic sign; (ii) we print the interlaced image on photo-quality paper and stick it on the back of a commercially available lenticular lens. We have used a free software called ``SuperFlip'' available online to interlace the chosen images \cite{freesupe78:online}.

\section{Experimental goals and setup}
In this section, we describe our experimental goals and setup, which motivate the results in the subsequent sections.

\subsection{Experimental goals}\label{subsec: eval_goals}
It is well-known that \advtraffic adversarial examples are extremely effective in the virtual setting. With our experiments, we show that \signembedding attacks are effective in both virtual and real-world settings. In particular, we seek to demonstrate with an appropriate pipeline for the generation, evaluation and selection of adversarial examples, an adversary can achieve high real-world attack success rates. In particular, we answer the following questions in our experiments:
\begin{enumerate}
	\item How effective are \signembedding attacks starting from high-resolution real-world images on standard classifiers? Answer: Section \ref{sec: white-box}
	\item How does an adversary generate and evaluate adversarial examples for a real-world setting? Answer: Section \ref{subsec: sign_embedding_real}
	\item Can \signembedding attacks break existing state-of-the-art defenses for neural networks? Answer: Section \ref{subsec: defense}
	\item Are transferability-based black-box attacks effective in real-world settings? Answer: Section \ref{sec: black-box}
	\item Can the \lentprinting attack fool classifiers? Answer: Section \ref{subsec: lent_eval}
\end{enumerate}

Overall, we demonstrate that with our pipeline, it is a low overhead procedure for an adversary to test both \advtraffic and \signembedding adversarial examples and subsequently use these to carry out effective real-world attacks.

\subsection{Experimental setup}\label{subsec: exp_setup}
For our experiments we used a GPU cluster of 8 NVIDIA Tesla P100 GPUs with 16GB of memory each, running with no GPU parallelization. A summary of the datasets and classifiers used is given in Table \ref{tab: eval_summary}. The overall pipeline was described earlier in Figure \ref{fig: detect}. Dataset details are in Appendix \ref{app: data_details}, detector details in Appendix \ref{app: detector} and classifier details in Appendix \ref{app: class_details}.

\begin{table*}
 \centering
 \resizebox{0.99\textwidth}{!}{
\begin{tabular}{c|c|c|c|c}
 \toprule
 Attacks & Virtual attack success (VAS) & Simulated physical attack success (SPAS) & Avg. norm ($L_1$) & Avg. confidence \\
 \midrule
 \advtraffic (\auxdata) &  54.34 \% &  36.65 \% & 37.71 & 0.9721  \\
 \signembedding (\adsign) & 85.71\% & 65.07\% & 34.89 & 0.9753 \\
 \signembedding (\customsign) & 29.44\% & 18.72\% & N.A. & 0.9508 \\ 
 \bottomrule
\end{tabular}
}
\caption{White-box attack success rates for \advtraffic and \signembedding attacks on the \mltscl
(virtual setting).}
\label{tab: compare_to_cw}
\vspace{-15pt}
\end{table*}

\subsubsection{Metrics} \label{subsubsec: eval_metrics}
Our experiments only consider \emph{targeted attacks}.

\noindent \textbf{Virtual attack success (VAS):} This is the standard evaluation metric used in previous work \cite{Carlini16,liu2016delving,bhagoji2017exploring} where the success rate is measured as the fraction of adversarial examples that are classified as their target.

\noindent \textbf{Simulated physical attack success (SPAS):} Manually printing out and evaluating the effectiveness of adversarial examples in a real-world setting is prohibitively expensive. In light of this, we propose a method to evaluate how physically robust adversarial examples are likely to be by simulating varying conditions in which the images of them might be taken. The physical conditions are virtually simulated by a combination of the randomized image transformations described earlier (Section \ref{subsubsec:transforms}). \textit{10 randomized, composite transformations} (brightness, perspective and re-sizing) are applied to each adversarial and original sample, and the transformed images are directly fed into the target classifier to determine the predicted labels. The \emph{simulated physical attack success} (SPAS) is then the fraction of these transformed images that are classified in the adversary's desired target class divided by the total number of transformed images. 


\noindent \textbf{Perceptibility and confidence:} The perceptibility of the perturbations is measured by computing the average $L_1$ norm of the perturbation for all adversarial examples.  We compute and report the average confidence of the target class over all \textit{successful} adversarial examples.

\noindent \textbf{Efficiency:} The slowest step in the generation of each adversarial example is running the optimization from Eq. \eqref{eq: opt_problem}. Each example takes about 60s to generate on the hardware we use. We stop the optimization run after 3000 steps for each sample since the change in loss between subsequent runs is vanishingly small.

\noindent \textbf{Remark (virtual and physical settings):} All results reported in the \emph{virtual} setting involve benign and adversarial examples that remain in a digital format throughout. The results for \emph{simulated physical attacks} (SPA) also pertain to images of this nature. The evaluation setup and metrics used for \emph{real-world attacks} are described in Section \ref{subsec: sign_embedding_real}.


\section{White-box attacks} \label{sec: white-box}
In this section we present results for both virtual and real-world attacks, in the white-box setting for adversarial examples. We also evaluate the effectiveness of our attacks against classifiers trained using adversarial training and its variants \cite{goodfellow2014explaining,madry_towards_2017}. 

\textbf{Attack pipeline in practice:} Equation \ref{eq: opt_problem} has a number of parameters such as batch-size $B$, $K$ to adjust the confidence of the generated adversarial examples, $c$ to control the trade-off between distance and loss etc. We tuned these parameters as described in Appendix \ref{app: param_tune} to achieve high SPAS rates. Our attack pipeline, described in Figure \ref{fig: attack} up-samples the generated perturbations to match the dimensions of the original image. When testing, we again down-sample these images so that they can be classified.

\subsection{Virtual \signembedding attacks} \label{subsec: sign_embedding_virtual}
We first evaluate the effectiveness of the \signembedding attack on high-resolution images. This allows us to pick the adversarial examples that work well in this setting for the real-world experiments. We experiment with 2 types of \signembedding images: \adsign and \customsign.

\subsubsection{\adsign attack} 
In this attack, we modify high-resolution images of innocuous signs and logos such that they are classified in the desired target class.  Each of the 7 original logos is passed through our attack pipeline to create 20 different \adsign adversarial examples, each meant to be classified as a different, randomly chosen target traffic signs, giving a total of 140 adversarial examples. This attack achieves an impressive \textbf{VAS of 85.71\%} and \textbf{SPAS of 65.07\%} with an average confidence of 0.975 as reported in Table \ref{tab: compare_to_cw}.  While the adversarial perturbation on the logos significantly affects the classification results, it is generally not very visible to humans and usually blends in with the detail of the logos. Some successful \adsign attacks are displayed in Figures \ref{fig: logo} and \ref{fig: logo_adv} (Appendix).

\subsubsection{\customsign attack} 
3 circular blank signs of the colors blue, orange, and white are drawn on the computer and used as the original images for the \customsign attacks. Each blank sign is matched with six different masks (a cricle, a heart, a star, a cross, the phrase "Hello, World!" and a combination of these shapes). Each of the total 18 pairs of a blank sign and a mask is used to generate 10 \customsign attacks which would be classified as 10 randomly chosen target traffic signs. In total, 180 \customsign attacks are created for the virtual evaluation. Some \customsign attacks are shown in Figures \ref{fig: custom} and \ref{fig: custom_adv} (Appendix). The attack produces adversarial signs that contain the shape of the mask filled with various colors. In Figure \ref{fig: custom}, we pick some of the signs whose mask is filled well so the text or the shape is clearly visible. Some of the attacks do not fill up the entire mask resulting in incomplete shapes. This attack is considerably different from the \adsign and the \advtraffic attacks as the optimization constraint is moved from the norm of the perturbation to its location instead. The adversary is allowed to add arbitrary perturbations as long as they are within the mask, hence the average $L_1$ norm is not applicable. This new constraint seems to be more strict than the previous one, causing the attack success rate to drop as shown in Table \ref{tab: compare_to_cw}.

\noindent \textbf{Main takeaway:} Both the \signembedding \adsign and \customsign attacks are feasible in the virtual white-box setting with high-confidence classification for the successful adversarial examples. Their non-trivial SPAS enables the adversary to pick effective adversarial examples from this setting for real-world attacks.

\subsection{Virtual \advtraffic attacks} \label{subsec: adv_traffic_sign}
We use high-resolution images from the \auxdata to generate \advtraffic attacks for the physical setting since the GTSRB test images are too low-resolution for large-scale printing. Both classifiers achieve 100\% classification accuracy on these images in the benign setting. Similar to the \adsign attack experiment, each of the 22 images in the \auxdata is used to generate 20 \advtraffic attacks for 20 randomly chosen target traffic signs. Therefore, one experiment contains 440 attacks in total. From Table \ref{tab: compare_to_cw}, we can see that attacks using these samples are feasible, confirming previous work.

\subsection{Real-world attacks (\signembedding and \advtraffic)} \label{subsec: sign_embedding_real}

\begin{figure*}
	\centering
	\includegraphics[width=0.9\textwidth]{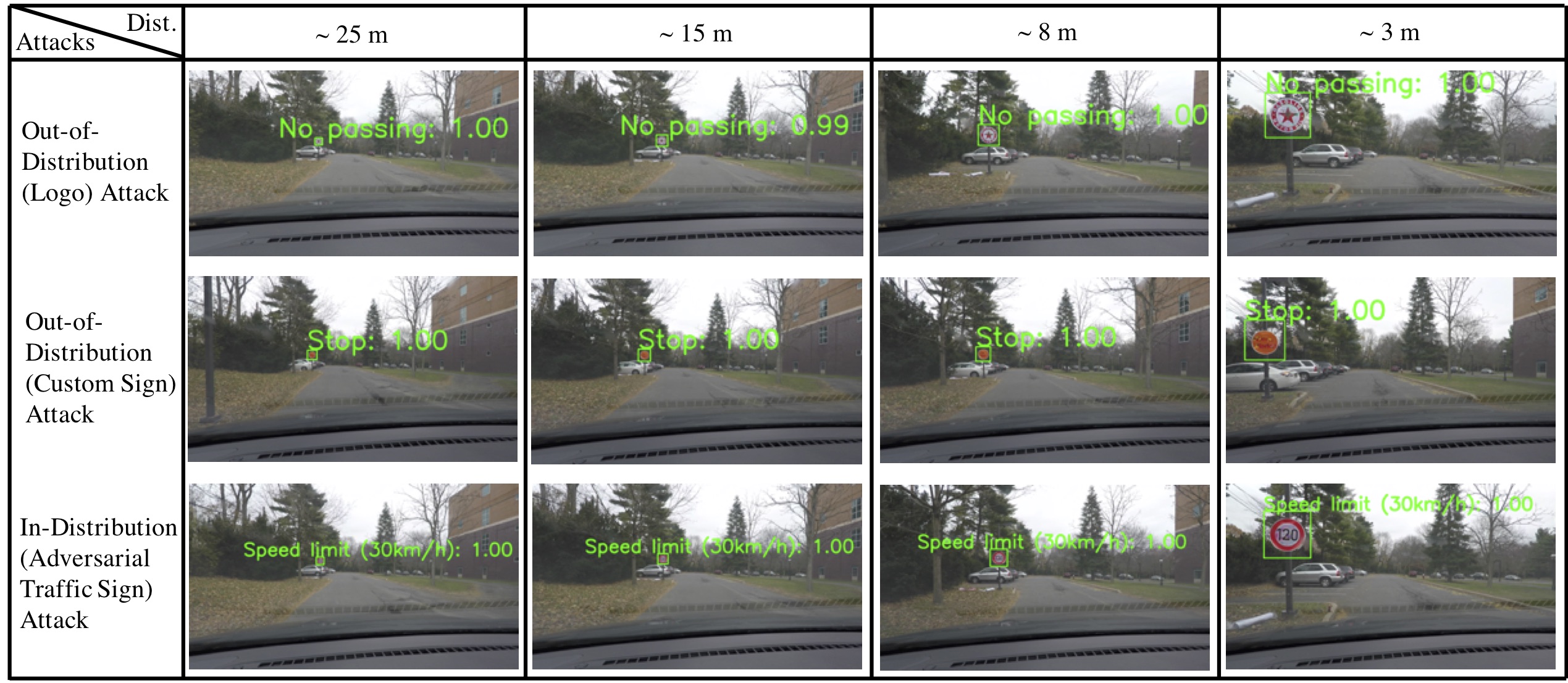}
	\caption{\textbf{Classification outcomes for adversarial examples in the drive-by test.} All combinations of distances and attacks lead to the desired classification outcome with high confidence. Even frames captured at around 25 metres from the sign lead to high confidence targeted misclassification.}
	\label{fig: real-world_result}
	\vspace{-10pt}
\end{figure*}

To carry out a real-world attack, we first chose 2 adversarial examples from each of the Logo, Custom Sign and Auxiliary traffic sign datasets that performed well under the Simulated Physical Attack (i.e. had consistent high confidence classification). Each of these examples was re-sized to $30\times30$ inches and printed on a high-quality poster. The printed signs are stuck on a pole at a height of 2 meters from the ground on the left side of the road with respect to a car driving towards the pole. A GoPro HERO5 was mounted behind the car's windshield to take videos of $2704\times1520$ pixels at 30 frames per second. Starting around 25 meters from the sign, the car approached it with an approximate speed of 16kph. The traffic sign recognition pipeline only detects and classifies once every 5 frames to reduce the processing time and redundancy. 

\noindent \textbf{Real-world attack metric:} To evaluate real-world attacks, we use a different metric since the number of adversarial examples used is smaller. In this section, the attack success rate reported is computed by counting the number of frames in which a sign was detected in the input frame and classified as the target class and dividing this number by the total number of frames in which there was a detection. In other words, we can express this as:
\begin{equation}
\text{Drive-by attack success} = \frac{\text{No. of frames sign is misclassified}}{\text{No. of frames sign is detected}}
\end{equation}

\begin{table}[t]
	\centering
	\resizebox{0.49\textwidth}{!}{
		\begin{tabular}{m{3.5cm}|m{2.5cm}|m{2.5cm}} 
			\toprule
			Attacks & White-box (Avg. confidence) & Black-box (Avg. confidence) \\
			\midrule
			\advtraffic (\auxdata) & 92.82\% (0.9632) & 96.68\% (0.9256) \\
			\signembedding (\adsign) & 52.50\% (0.9524) & 32.73\% (0.9172) \\
			\signembedding (\customsign) & 96.51\% (0.9476) & 97.71\% (0.9161) \\
			\bottomrule
		\end{tabular}
	}
	\caption{Real-world attack success rates against the \mltscl in the white-box setting (average of 2 signs for each attack) and on \cnn in the black-box setting (best performing sign samples transferred from \mltscl).}
	\label{tab: realworld_results}
	\vspace{-25pt}
\end{table}

\noindent \textbf{Real-world attack success:} To demonstrate the effectiveness of our adversarial examples in the real world, we carried out \emph{drive-by tests} (shown in Figure \ref{fig: real-world_result}) on 2 samples from each of our attacks (\advtraffic, \adsign, and \customsign). Each of the drive-by attack success rates reported in Table \ref{tab: realworld_results} is an average of three runs. We hypothesize that the lower attack success rate for the \adsign attack in the drive-by setting occurs because the particular instance that was chosen had a more difficult target class. However, for frames classified as the target, the confidence is high. The source-target pairing for the \advtraffic attack was two speed limit signs, which could have contributed to the high attack success.

\noindent \textbf{Main takeaway.} In the drive-by tests, both the \signembedding and \advtraffic adversarial examples can achieve attack success rates in excess of 90\%.

\subsection{Attacking defended models} \label{subsec: defense}
In this section we examine the effectiveness of the \signembedding and \advtraffic attacks against state-of-the-art defenses based on the concept of \emph{adversarial training} \cite{goodfellow2014explaining}. In this paper, we are the first to analyze the effectiveness of adversarial training against physically robust adversarial examples. We note that this defense mechanism is completely ineffective against lenticular printing based attacks.  

\subsubsection{Adversarial training} This defense modifies the loss function of neural networks to increase their robustness against adversarial examples. The \emph{training loss} is modified as follows:
\begin{align}
\label{eq: adv_training}
\ell^{\text{adv}}_f(\bfx,y) = \alpha \ell_f(\bfx,y) + (1 - \alpha) \ell_f(\xad,y),
\end{align}
where $\alpha \in [0,1]$ controls the adversarial component of the loss and $\xad$ is an adversarial example. We trained \mltscl using the loss function above with $\alpha=0.5$ and using Fast Gradient Sign adversarial examples \cite{goodfellow2014explaining} with an $\epsilon=0.3$. The model was trained for 15 epochs and has an accuracy of \textbf{96.37\%} on the GTSRB validation set. We do not use the adversarial examples generated using Equation \eqref{eq: opt_problem} as each sample takes prohibitively long to generate for the purposes of training. Further details are in Appendix \ref{app: adv_training}.

\begin{table}[t]
	\centering
	\resizebox{0.49\textwidth}{!}{
		\begin{tabular}{m{3.5cm}|m{1cm}|m{1cm}|m{1.5cm}|m{1.5cm}} 
			\toprule
			Attacks & \sucrate & \sucratephy & Avg. confidence & Avg. norm ($L_1$) \\
			\midrule
			\advtraffic (\auxdata) & 2.53\% & 2.47\% & 0.9358 & 38.92\\
			\signembedding (\adsign) & 11.42\% & 7.42\% & 0.8054 & 36.12 \\
			\signembedding (\customsign) & 6.67\% & 5.77\% & 0.9957 & N.A. \\
			\bottomrule
		\end{tabular}
	}
	\caption{\textbf{Attack success rates and deterioration rate on \textit{adversarially trained} \mltscl for \signembedding and \advtraffic attacks in the virtual white-box setting.}}
	\label{tab: results_adv_training}
	\vspace{-15pt}
\end{table}

\subsubsection{Effect of \signembedding attacks on adversarially trained models}
The virtual white-box attack results given in Table \ref{tab: results_adv_training} demonstrate that adversarial training is very effective as a defense against \advtraffic attacks since the VAS using \auxdata drops to 54.34\% to 2.53\% while the SPAS drops from 36.65\% to 2.47\%. On the other hand, the VAS and SPAS for the \signembedding \adsign attack remain at 11.42\% and 7.42\% respectively. Thus, the \adsign attack is about $3-4 \times $ more effective than the \advtraffic attack. The \customsign attack is also more effective that the \advtraffic attack, indicating that \signembedding attacks are more effective against these defenses.

\textbf{Main takeaway:} Adversarial training cannot defend against \lentprinting and \signembedding attacks are more effective than previously examined \advtraffic attacks. Thus, we have introduced new attack vectors against these state-of-the-art defenses.

\begin{table}[t]
	\centering
	\resizebox{0.49\textwidth}{!}{
		\begin{tabular}{m{3.5cm}|m{2.2cm}|m{2.3cm}|m{2.5cm}} 
			\toprule
			Attacks & Black-box \sucrate & Black-box \sucratephy & Avg. confidence\\
			\midrule
			\advtraffic (\auxdata) & 7.14\% & 6.08\% & 0.9273 \\
			\signembedding (\adsign) & 14.28\% & 12.57\% & 0.8717 \\
			\signembedding (\customsign) & 3.33\% & 6.00\% & 0.7193 \\
			\bottomrule
		\end{tabular}
	}
	\caption{ \textbf{Black-box attack success rates for \advtraffic and \signembedding attacks.} Adversarial examples generated for \mltscl are tested on \cnn.}
	\label{tab: blackbox_results}
	\vspace{-20pt}
\end{table}

\section{Black-box attacks are possible} \label{sec: black-box}
In the black-box attack setting, we use adversarial examples generated for the \mltscl and test their attack success rate on \cnn in both virtual and real-world settings. The recognition pipeline is kept the same, but the classifier is changed. The reason this attack is expected to work is the well-known phenomenon of \emph{transferability} \cite{papernot2016transferability,papernot2016practical,liu2016delving}, where adversarial examples generated for one classifier remain adversarial for another.  Since both classifiers are trained on the same dataset, we are implicitly assuming that the adversary has access to the training data. This is a common assumption in previous work on transferability-based black-box attacks.

\noindent \textbf{Virtual attacks:} Table \ref{tab: blackbox_results} shows adversarial success rate in a black-box setting, demonstrating the transferability of adversarial examples. While there is a significant drop in attack success rates, both the VAS and SPAS for the \signembedding \adsign attacks remain at non-negligible rates of above 10\% indicating that even an attacker with very limited knowledge of the target classifier can carry out successful attacks.

\noindent \textbf{Real-world attacks:} We use the black-box attack pipeline to evaluate the same videos captured in the drive-by tests (Table \ref{tab: realworld_results}). With the same set of signs that perform best in the white-box case, the \signembedding \customsign and the \advtraffic attacks both achieve very attack high success rates of \textbf{97.71\%} and \textbf{96.68\%} respectively, comparable to the white-box setting. One of the \adsign attacks achieves an attack success rate of 32.73\% in the black-box case. Note that the higher attack success rates for the black-box setting arise from the fact that we just report numbers for the best-performing adversarial sign and not an average over two signs.
\noindent \textbf{Main takeaway:} Black-box attack success rates in excess of 90\% can be achieved in real-world settings with appropriately chosen adversarial examples. This represents a significant threat against traffic sign recognition systems since no system knowledge is required for these.


\subsection{Lenticular printing attacks} \label{subsec: lent_eval}
In these experiments, we aim to show that the difference in viewing angle between the camera on an AV and the human controller or passenger can lead to an attack, since the human will see an appropriate sign, while the camera will see the targeted sign, which could be dangerous in that environment. To simulate what the human/camera will see, we take pictures from different heights showing the sign changing. We emphasize that the adversarial nature of these examples stems purely from the viewing angle, and the human and the camera would recognize the sign as the same if they viewed it from the same angle.

\begin{figure}[t]
	\centering
	\subfloat[Camera height of $1.5$ metres. Signs classified as `Speed limit 60', `Turn left' and `Turn right' respectively, with a confidence of 1.0.]{\includegraphics[width=0.25\linewidth]{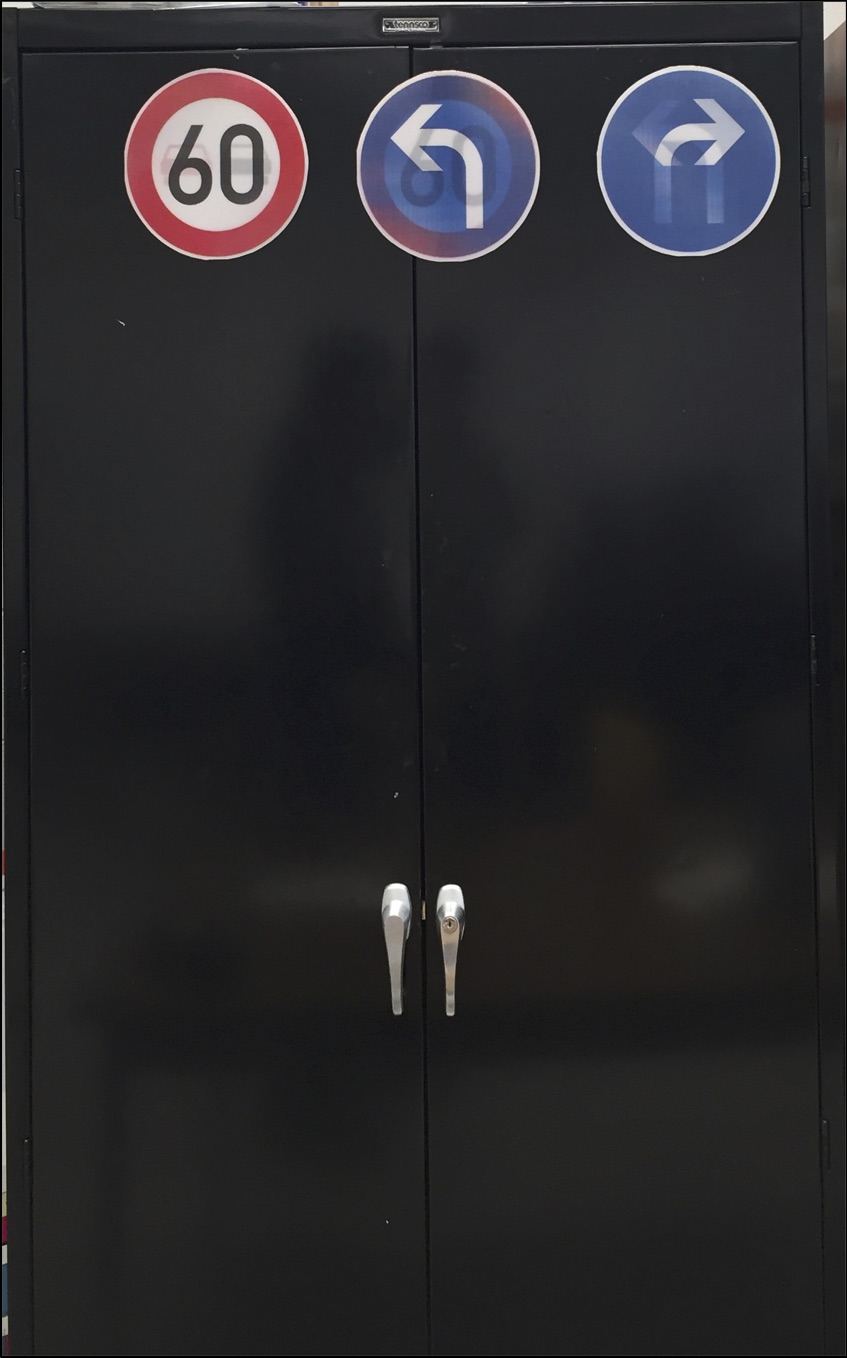}\label{fig: len_1}} \qquad
	\subfloat[Camera height of $0.8$ metres. Signs classified as `No passing', `Speed limit 60' and `Turn left' respectively, with a confidence of 1.0. ]{\includegraphics[width=0.25\linewidth]{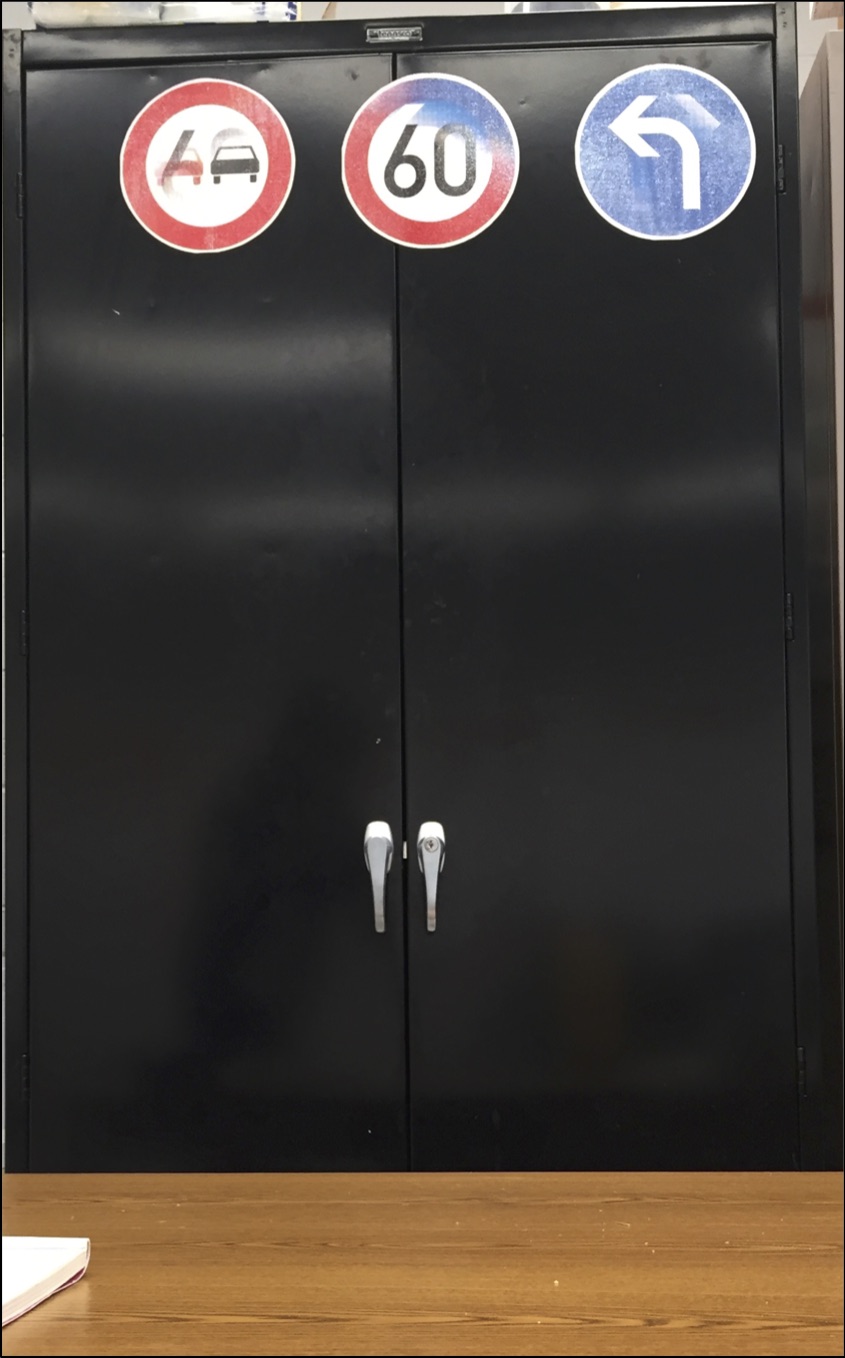}\label{fig: len_2}} \qquad
	\subfloat[Camera height of $0.5$ metres. Signs classified as `No passing', `Speed limit 60' and `Turn left' respectively, with a confidence of 1.0.]{\includegraphics[width=0.25\linewidth]{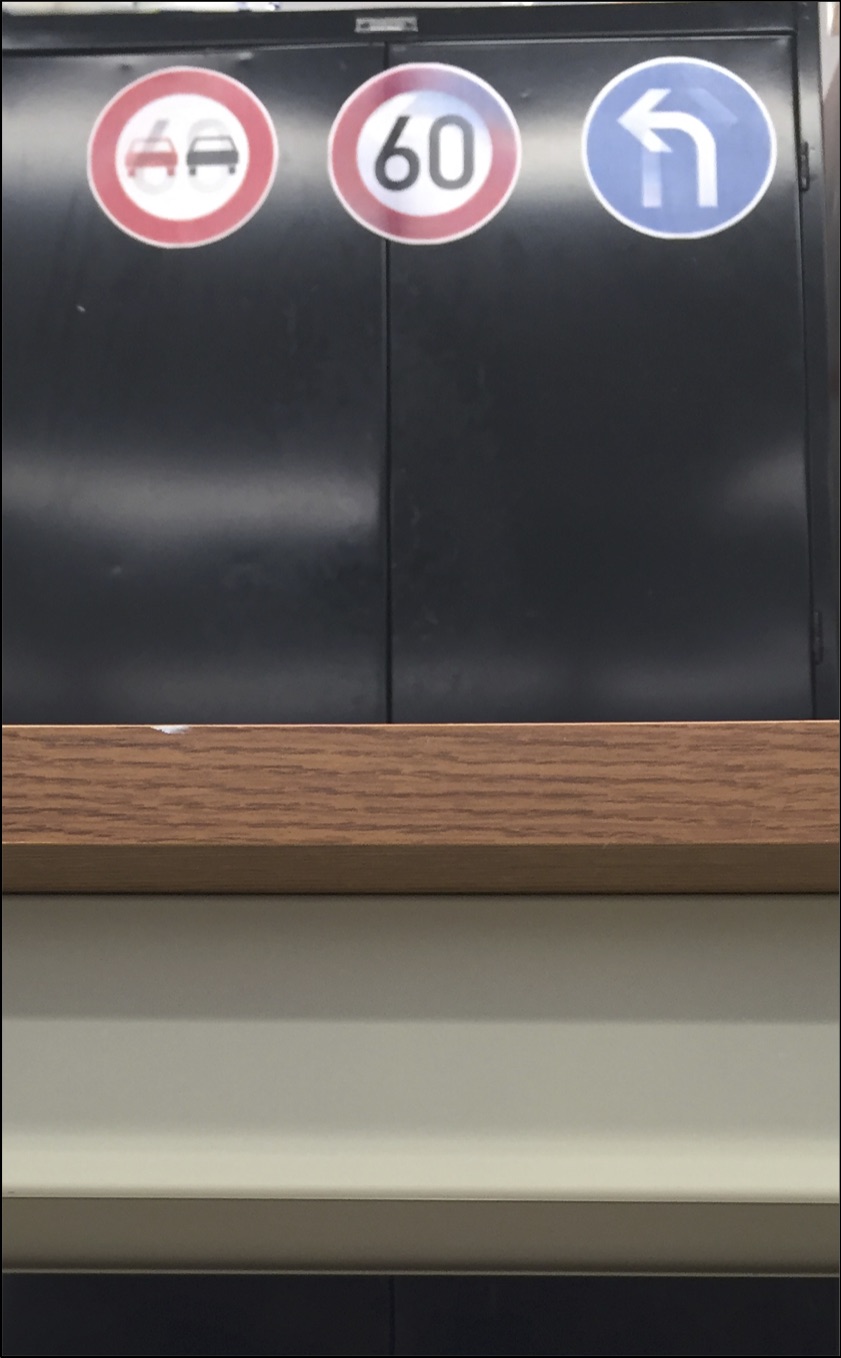}\label{fig: len_3}}
    \vspace{-10pt}
	\caption{\textbf{Proof-of-concept implementation of the \lentprinting attack.} These images show that if the camera used for the traffic sign recognition module of an AV is at a different height from the human controller, then the \lentprinting attack can fool the classifier, while appearing to be correct to the human.}
	\label{fig: lenticular_exp}
	\vspace{-15pt}
\end{figure}

In our experiments with the \lentprinting attack, we stuck three different signs on an indoor surface (Figure \ref{fig: lenticular_exp}). These signs flip between `Speed limit 60' and `No Passing', 'Turn left' and 'Speed limit 60', and 'Turn left' and 'Turn right' respectively. We take pictures of these signs from three different observation angles ($\theta_1=8$, $\theta_2=25$, and $\theta_3=33$ degrees) using a standard mobile phone camera at a distance of $1.9$ metres from the signs. The camera is held parallel to the plane of the cupboard. In all cases, when passed through our traffic sign recognition pipeline, each sign classified was classified as the appropriate one for that viewing angle with high confidence. 
\textbf{Validating theory:} Based on the theoretical discussion in Section \ref{sec:lent}, we expect that the camera observes two images from two different angles if the difference between these two observation angles is between $12.5$ and $25$ degrees. This matches our observations demonstrated in Figure \ref{fig: lenticular_exp}: (1) $25 \geq |\theta_1-\theta_2| = 17 \geq 12.5$ and $25 \geq |\theta_1-\theta_3|=25 \geq 12.5$, indicating that the first image should look different from other images, and (2) $|\theta_2-\theta_3|=8 \leq 12.5$, suggesting that the second and third images should look similar.  

\textbf{How can the attacker ensure that the driver cannot see different images as the car approaches the sign?} The attacker can readily measure the range of the observation angle of the driver and the camera given the height of the driver's eyes, the height of the camera, and height of the sign. The attacker can then create and install the sign, while considering the theoretical discussion in Section \ref{sec:lent}. A straightforward estimation of the height of the driver's eyes and the height of the camera can be made based on the height of the target vehicle. For simplicity, the attacker can install the sign on a flat road at the height of the driver's eyes. In this case the observation angle of the driver remains fixed, in particular, $\theta_1=0$.

\section{Limitations and Future Work}\label{sec: discussion}
In this section, we discuss some limitations of our approaches for both attacks and defenses and outline possibilities for future work to overcome these limitations.

\noindent \textbf{Adversarial example detectors as a countermeasure:} While detection based defenses such as Feature Squeezing \cite{xu2017feature} and MagNet \cite{meng2017magnet} are ineffective against \advtraffic white-box attacks \cite{carlini2017magnet}, it is an open question as to how they will perform against \signembedding attacks. We plan to explore detection-based defenses such as these in future work.

\noindent \textbf{Fooling synthesis of sensor inputs:}
The computer vision subsystem of an AV, while critical, is not the only actor in the decision making process when confronted with a new situation. A number of other sensors such as LIDAR, GPS, radar etc. also provide inputs which are then synthesized to come up with a decision \cite{mujica2014scalable}. While the computer vision subsystem is the only one able to recognize a traffic sign, the other sensors may be able to indicate that the sign recognized is incompatible with their inputs. This consideration is out of the scope of the current work, but we plan to explore simultaneous attacks on these varied subsystems in future work.


\noindent \textbf{Choice of norm:}
In this paper, we chose the $L_1$ norm to measure the visibility of adversarial perturbations as done in several previous works on virtual attacks \cite{fawzi2015analysis,Carlini16, chen2017ead}. However, it is still an open research question if this is the best choice of distance function to constrain adversarial perturbations. Previous work on generating adversarial examples has explored the appropriateness of other commonly used Euclidean norms such as $L_{\infty}$ \cite{goodfellow2014explaining} and $L_2$ \cite{Carlini16,papernot2016transferability} as well. Especially in the context of future work on generating physically realizable adversarial examples, we encourage further work on conducting user studies to determine the most appropriate proxy for measuring human perceptibility of adversarial perturbations.

\section{Related work}\label{sec: rel_work}
\noindent \textbf{Virtual white-box attacks:} Evasion attacks (test phase) have been proposed for Support Vector Machines \cite{biggio2014security,moosavi2015deepfool}, random forests \cite{kantchelian2016evasion} and neural networks \cite{szegedy2013intriguing,goodfellow2014explaining,papernot2016limitations,Carlini16,moosavi2015deepfool,moosavi2016universal}. Nguyen et al. \cite{nguyen2015deep} generate adversarial examples starting from random noise restricted to the virtual white-box setting. Poisoning attacks (training phase) have also been proposed for a variety of classifiers and generative models \cite{biggio2012poisoning,rubinstein2009stealthy,poison-lasso,poison-lda,poison-autoregressive}. Attacks have also been proposed on policies for reinforcement learning \cite{huang2017adversarial}, models with structured prediction outputs \cite{cisse2017houdini}, semantic segmentation models \cite{arnab2017robustness,fischer2017adversarial} and neural network based detectors \cite{xie2017adversarial}.

\noindent \textbf{Virtual black-box attacks:} The phenomenon of transferability \cite{szegedy2013intriguing,papernot2016transferability} has been used to carry out black-box attacks on real-world models \cite{papernot2016practical,liu2016delving}. Powerful query-based black-box attacks have also been demonstrated \cite{nelson2010near,nelson2012query,lowd2005adversarial,chen2017zoo,bhagoji2017exploring,brendel2017decision}.

\noindent \textbf{Real-world attacks:} Kurakin et al. \cite{kurakin2016adversarial} were the first to explore real-world adversarial examples. They printed out adversarial examples generated using the Fast Gradient Sign attack and passed them through a camera to determine if the resulting image was still adversarial. This attack was restricted to the white-box setting and the effect of varying physical conditions was not taken into account. Sharif et al. \cite{Sharif16} investigated how face recognition systems could be fooled by having a subject wear eyeglasses with adversarial perturbations printed on them. While their attack results were encouraging, they did not rigorously account for varying physical conditions and only took multiple pictures of a subject's face to increase the robustness of the attack. Further, they only tested their black-box attacks in a virtual setting with query access to the target model. Petit et al. \cite{petit2015remote} examine the susceptibility of the LIDAR and camera sensors in an autonomous car but their attacks are unable to cause targeted misclassification. Lu et al. \cite{lu2017adversarial} and Chen et al. \cite{chen2018robust} attempted physical-world attacks on R-CNN based traffic sign detectors \cite{Ren15frcnn}. However, large perturbations were needed in both cases. We plan to explore \signembedding attacks on these detectors in future work. Comparisons with the related work of Athlaye et al. \cite{Athalye17} and Evtimov et al. \cite{Evtimov17} have been elucidated in Section \ref{sec: intro}.

\section{Conclusion}\label{sec: conclusion}
In this paper, we have demonstrated a wide range of attacks on
traffic sign recognition systems, which have severe consequences
for self-driving cars. \signembedding attacks allow an adversary to convert any sign or logo into into a targeted adversarial example. The \lentprinting attack moves beyond the paradigm of adversarial examples to create images that look different from varying heights, allowing an adversary to stealthily embed a potentially dangerous traffic sign into an innocuous one, with no access to the internals of the classifier. We demonstrated the effectiveness of our attacks in both virtual and real-world settings. We are the first to carry out black-box attacks in a real-world setting as well as to evaluate possible countermeasures against physical realizations of robust adversarial examples. We hope our discussion of future research directions encourages further exploration into securing physically deployed machine learning systems. 

\bibliographystyle{unsrt}
\bibliography{ref.bib}

\appendix

\section{Appendix: Dataset details} \label{app: data_details}
\begin{figure*}[t]
	\centering
	\includegraphics[width=0.8\textwidth]{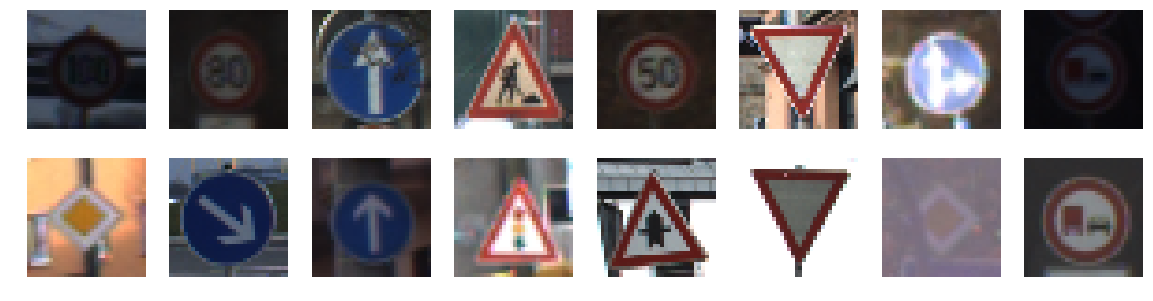}
	\caption{Samples from the GTSRB dataset}
	\label{fig: gtsrb}
\end{figure*}

We use the German Traffic Sign Recognition Benchmark (GTSRB) \cite{Stallkamp2012}, a widely-used standard dataset for traffic sign recognition, to train and test both classifiers \footnote{We chose GTSRB over the LISA Traffic Sign Dataset \cite{LISA} as GTSRB offers a much larger number of samples leading to better generalized classifiers. }. It contains 51,839 images of 43 types of traffic signs in total. Each image is a $32 \times 32 $ RGB image with the pixels scaled to lie in $[0,1]$. Sample images are shown in Figure \ref{fig: gtsrb}.The 39,209 training samples are augmented to 86,000 (2,000 for each class) using perspective transformation and flipping signs that are invariant to flipping (i.e. "no entry") or that are interpreted as a different class after flipped (i.e. "turn right" to "turn left").

\begin{figure*}[t]
	\centering
	\includegraphics[width=0.8\textwidth]{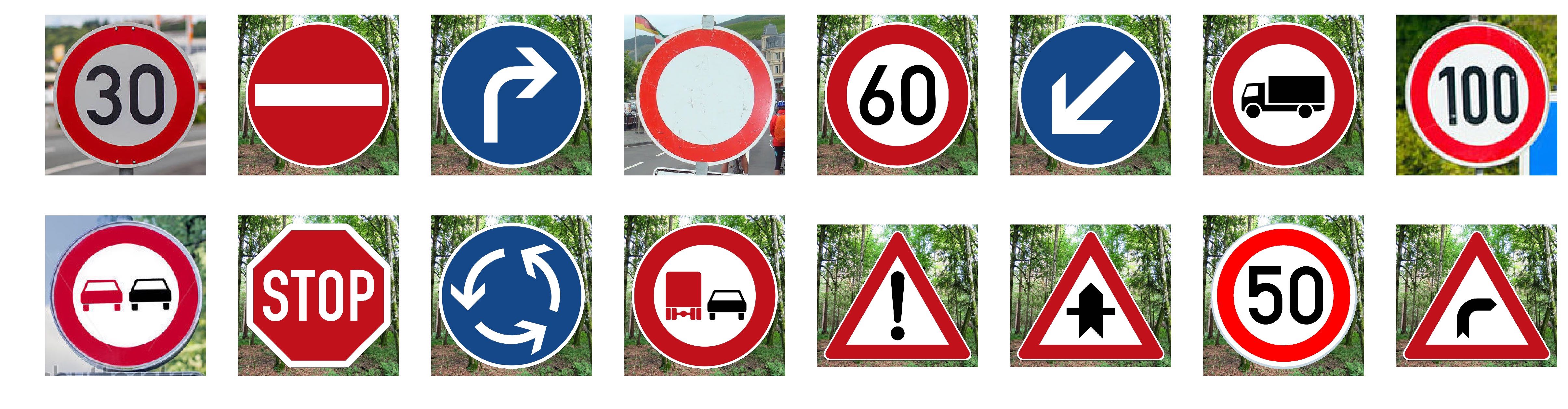}
	\caption{\textbf{The \auxdata}. Some high-resolution images sampled from the \auxdata. The \auxdata is a combination of real traffic sign photos and computer-generated drawings in front of a simple background.}
	\label{fig: auxdatasign}
\end{figure*}

\noindent \textbf{Auxiliary high-resolution dataset:} To create fake traffic signs that look realistic, we used 22 high-resolution traffic sign images to use as original images for the \advtraffic attack, shown in Figure \ref{fig: auxdatasign}. They are a mixture of real photographs and computer-generated drawings on an arbitrary background image.

\begin{figure*}[t]
	\centering
	\includegraphics[width=0.8\textwidth]{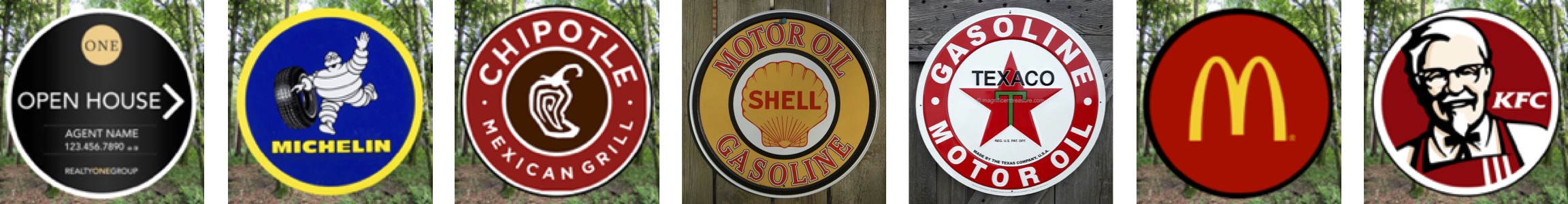}
	\caption{Logo dataset. High-resolution images of common logos.}
	\label{fig: logosign}
\end{figure*}

\noindent \textbf{Logo dataset:} To create realistic-looking \signembedding \adsign adversarial examples, we used 7 high-resolution Logo images on arbitrary backgrounds as shown in Figure \ref{fig: logosign}.

\begin{figure*}[t]
	\centering
	\includegraphics[width=0.8\textwidth]{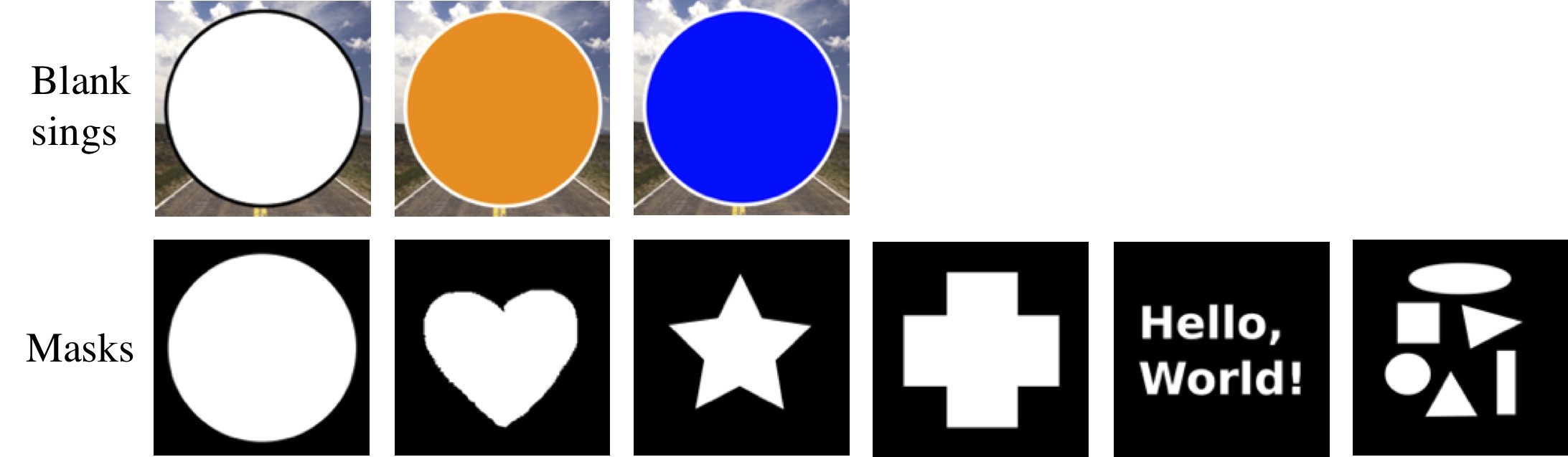}
	\caption{Custom Sign dataset. Blank signs with 6 different masks.}
	\label{fig: customsign}
\end{figure*}

\noindent \textbf{Custom Sign dataset:} To create realistic-looking \signembedding \customsign adversarial examples, we used 3 circular blank signs (orange, blue and white) in combination with 6 masks on arbitrary backgrounds as shown in Figure \ref{fig: customsign}.

\section{Appendix: Detector details} \label{app: detector}
For simplicity and without loss of generality, we design our detector to only locate circular signs on images using a well-known shape-based object detection technique in computer vision, Hough transform \cite{HoughCir83:online}. Triangular signs can be detected by a similar method described in \cite{Yakimov2015Hough}. In detail, our pipeline takes a video frame as an input and smooths it with a Gaussian filter to remove random noise. The processed images are passed through a Canny edge detector and then a circle Hough transform which outputs coordinates of the center and radii of the circles detected. The outputs are used to determine a square bounding box around the detected signs. The section of the input frame inside the bounding boxes are cropped out of their original \textit{unprocessed} image and resized to $32\times32$ pixels which is the input size of the neural network classifier.

\section{Appendix: Classifier details} \label{app: class_details}
Our classifier in the \emph{white-box setting}, referred to as \mltscl is based on a multi-scale convolutional neural network (CNN) \cite{LeCun2011mltscl} with 3 convolutional layers and 2 fully connected layers. There is also an extra layer that pools and concatenates features from all convolutional layers, hence the name multi-scale. The training procedure is adpated from \cite{Traffics18:online}. The model is trained on data-augmented training set (86,000 samples, 2,000 for each class) generated by random perspective transformation and random brightness and color adjustment. There is no image preprocessing. The training takes 41 epochs with learning rate of 0.001. Dropout and weight regularization ($\lambda = 1\mathrm{e}{-3}$) are used to reduce overfitting. The accuracy of the model, which will be referred to as \mltscl, on the validation set is \textbf{98.50\%}. The model is defined in Keras \cite{chollet2015keras} using Tensorflow \cite{tensorflow2015-whitepaper} backend. The entire recognition pipeline (the detector combined with the classifier) is tested with the German Traffic Sign Detection Benchmark (GTSDB) \cite{Houben2013} where it achieves a reasonable mean average precision (mAP) \cite{Everingham2010} of \textbf{77.10\%} at the intersection-over-union (IoU) of 0.5. 

\cnn (used in the black-box attack setting) is a standard CNN (not multi-scale) with four convolutional layers and three dense layers. It is trained on the same dataset as \mltscl, and achieves an accuracy of \textbf{98.66\%} on the validation set of GTSRB. The mean average precision (mAP) of the entire pipeline with \cnn on the GTSDB dataset is \textbf{81.54\%}. In the black-box setting, the attacks are generated based on \mltscl which is assumed to be trained by the adversary as a substitute model for the true classifier in use, \cnn. All experiments in Section \ref{sec: black-box} use this black-box pipeline to evaluate the attacks. Both classifiers achieve a 100\% accuracy on the \auxdata, implying that they have generalized well to images of traffic signs.

\section{Appendix: Baseline results} \label{app: baseline results}
We compare the adversarial examples generated using our method to those generated by the optimization problem in Eq. \eqref{eq: cw_opti} (referred to as Vanilla Optimization) on a random subset of 1000 traffic signs chosen from the test data of the GTSRB dataset. These results are not comparable to the ones in the main text using high-resolution data since both the masking and re-sizing effects are absent from these.

The result shown in Table \ref{tab: baseline} demonstrates that our attack has a much lower deterioration rate (DR) compared to the Vanilla Optimization attacks. The Vanilla Optimization attack with the same parameters as our attack indicates that forcing perturbation to be large (having a large norm) can increase its robustness to transformation to a certain degree. However, our attack can achieve a much lower deterioration rate with a similar average norm emphasizing the substantial effect the addition of random transformations to the optimization problem has on improving the physical robustness of adversarial examples. As expected, both the VAS and SPAS are lower for the adversarially trained \mltscl and the black-box attack setting.

\begin{table*}
	\centering
	\resizebox{0.99\textwidth}{!}{
		\begin{tabular}{c|c|c|c|c|c|c}
			\toprule
			Models & Attacks & Virtual attack success (VAS) & Simulated physical attack success (SPAS) & Avg. norm ($L_1$) & Avg. confidence \\
			\midrule
			\multirow{2}{*}{\mltscl} & Baseline CW \cite{Carlini16} attack (GTSRB test data) & 97.91\% & 46.74\% & 30.45 & 0.9321 \\
			& \advtraffic (GTSRB test data)  & 99.07\% & 95.50\% & 31.43 & 0.9432 \\ \midrule
			\mltscl\textsubscript{adv} &\advtraffic (GTSRB test data) & 36.35\% & 27.52\% & 31.57 & 0.9428 \\ \midrule
			\multirow{1}{*}{\mltscl $\rightarrow$ \cnn} & \advtraffic (GTSRB test data) & 47.77\% & 38.08\% & 31.43 & 0.8838 \\
			\bottomrule
		\end{tabular}
	}
	\caption{White-box attack success rates for baseline and \advtraffic attacks on the GTSRB test data. The parameters for the Carlini-Wagner attack are modified in order to increase its Simulated Physical Attack Success rate.}
	\label{tab: baseline}
	\vspace{-10pt}
\end{table*}

\section{Appendix: Parameter tuning} \label{app: param_tune}
We use an additional metric, the \textit{Deterioration rate (DR)} in this section. This metric measures what fraction of the generated adversarial examples remain adversarial under simulated physical conditions: $\detrate = 1 - \frac{\sucratephy}{\sucrate}$. A higher deterioration rate implies that the adversarial examples are more likely to degrade under the transformations, and by extension in a real-world setting. 

The set of initial parameters was manually chosen to maximize the VAS on the \auxdata on \mltscl while maintaining a low average perturbation as measured in the $L_1$ norm. We use the $L_1$ norm since we found it to achieve the best trade-off between the various performance metrics such as VAS, SPAS and DR that we considered. Due to this trade-off, we chose not to use a grid search to fine the optimization parameters. Once we had a set of parameters that worked well, we changed each of the parameters individually to understand their effect on the attack performance metrics. Fist, we set the parameter $L$ to control the norm of the output perturbation. Depending on the type of norms used in the objective function, $L$ can be chosen roughly to any number larger than zero to force the algorithm to explore more solutions that cause targeted misclassification instead of finding those that merely minimize the norm. With $L$ chosen for each norm, we vary the constant $c$, norm $p$, the number of transformation $T$, and degree of randomness of the transformations. We use $\theta_1$ for perspective transformation, $\theta_2$ for brightness adjustment and $\theta_3$ for image resizing. We find that varying $\theta_3$ does not significantly affect the outputs and thus do not discuss it further. 

The baseline attack uses $L_1$ norm with the parameters $c=3, K=100, L=30, T=128, \theta_1=0.07, \theta_2=0.15$. Table \ref{tab: param_table} shows results from 13 experiments each of which vary specified parameters from the baseline. We use Adam optimizer to solve the optimization problem with learning rate (step size) of $0.02$ without decay. It must be noted that all four result columns of Table \ref{tab: param_table} must be viewed in conjunction, instead of only considering one column, because they represent the trade-offs of the attacks. For example, we can see that by increasing the degree of randomness ($\theta_1, \theta_2$), attack success rates generally increase, but the norms also become larger making the perturbation more noticeable. In particular, setting $\theta_1=0.15$ results in the second highest physical success rate and the lowest deterioration rate, but it also produces perturbation with the second largest norm. Similarly, setting $K$ to be larger encourages the optimization to look for more \textit{adversarial} solutions which also comes with the cost of a large perturbation. Increasing the number of transformations makes the attack more successful under transformation while reducing the success rate in the virtual setting slightly. In fact, using 512 transformations produces both high physical attack success rate and a small norm for the perturbation with only the expense of longer computation time. For an adversary who only needs a few adversarial signs, he or she can afford to use a large number of transformation or even run a search for optimal parameters for a specific setting. For the \customsign attack, since there is no constraint on the norm of the perturbation both $c$ and $L$ and are increased to obtain perturbations that fit within the mask and are adversarial.
Table \ref{tab: param_table} contains the results from tuning the various optimization parameters from Equation \ref{eq: opt_problem}.

\noindent \textbf{Main takeaway.} The number of tunable parameters in the optimization represents an advantage for the adversary since they can be tuned to achieve the desired trade-off between the different performance metrics. We chose a particular set of parameters that worked well for our evaluation setting, and clarified the changes in performance that occur when these are tweaked. 

\begin{table*}
	\centering
	\begin{tabular}{L{2.5cm}|L{4cm}|l|l|l|l} 
		\toprule
		Description & Parameters & \sucrate & \sucratephy & DR & Avg. norm ($L_1$) \\
		\midrule
		\multirowcell{1}{Chosen params}
		& $c=3, K=100, L=30, T=128, \theta_1=0.07,\theta_2=0.15$ & 54.34\% & 36.65\% & 32.55\% & 37.71 \\ 
		\midrule
		\multirowcell{3}{Perturbation\\norm}
		& $L_2$ norm ($c=0.2, L=2$) & 27.04\% & 15.09\% & 37.23\% & 76.74 \\
		& $L_2$ norm ($c=0.02, L=2$) & 14.03\% & 7.76\% & 44.69\% & 54.13 \\
		& $L_\infty$ norm ($c=5\mathrm{e}{-5}, L=0.1$) & 43.37\% & 24.41\% & 43.72\% & 162.76 \\
		\midrule
		\multirowcell{2}{Adversarial\\confidence}
		& $K=50$ & 48.21\% & 28.58\% & 40.71\% & 32.89 \\
		& $K=200$ & \textbf{59.69\%} & \textbf{50.91}\% & \textbf{14.71\%} & \textbf{\textcolor{red}{50.53}} \\
		\midrule
		\multirowcell{2}{\# of transformed\\samples}
		& $T=32$ & 54.59\% & 32.08\% & 41.23\% & 35.44 \\
		& $T=512$ & 46.43\% & 37.81\% & 18.57\% & 35.82 \\
		\midrule
		\multirowcell{6}{Degree of\\transformation\\($\theta_1$: perspective\\ transformation,\\$\theta_2$: brightness\\adjustment)}
		& $\theta_1=0, \theta_2=0$ & 31.89\% & 6.26\% & 80.37\% & 31.87 \\
		& $\theta_1=0.03$ & 44.13\% & 8.88\% & 79.88\% & 33.21 \\
		& $\theta_1=0.15$ & \textbf{51.79\%} & \textbf{44.88\%} & \textbf{13.34\%} & \textbf{\textcolor{red}{43.70}} \\
		& $\theta_2=0$ & 52.55\% & 34.56\% & 34.23\% & 36.61 \\
		& $\theta_2=0.075$ & 52.80\% & 35.68\% & 32.42\% & 36.94 \\
		& $\theta_2=0.30$ & 50.26\% & 39.34\% & 21.72\% & 39.40 \\
		\bottomrule
	\end{tabular}
	\caption{\textbf{Variation in attack success rates, deterioration rates and average norm of attack with different sets of optimization parameters.} Rows with numbers in \textbf{bold} represent parameter settings that achieve a good trade-off between the various performance metrics. However, both these rows have a higher average perturbation norm than the chosen set of parameters.}
	\label{tab: param_table}
	\vspace{-10pt}
\end{table*}

\section{Appendix: More Adversarial Examples} \label{app: more_adv}
\begin{figure*}[t]
	\centering
	\subfloat[Some samples of our In-Distribution (Adversarial Traffic Sign) attack, along with the labels they are classified as.]{\includegraphics[width=0.8\linewidth]{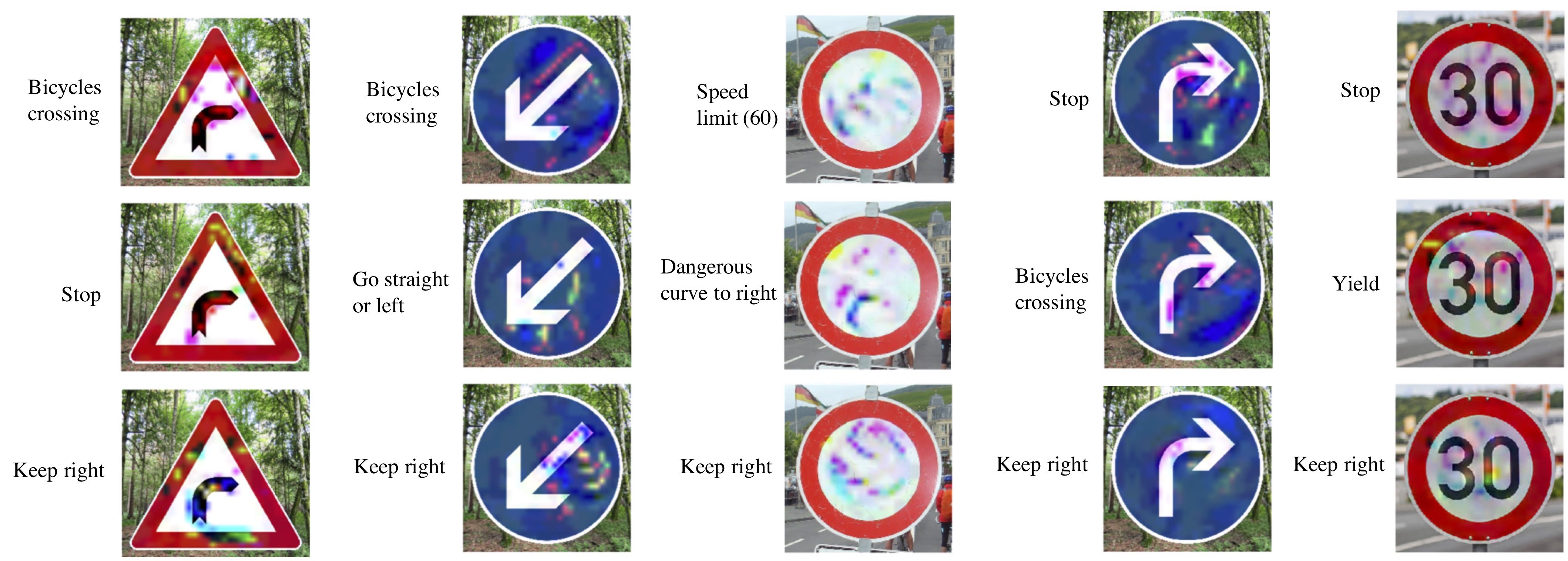}\label{fig: adv_sign_sample}} \\
	\subfloat[Some samples of our Out-of-Distribution (Logo) attack, along with the labels they are classified as.]{\includegraphics[width=0.54\linewidth]{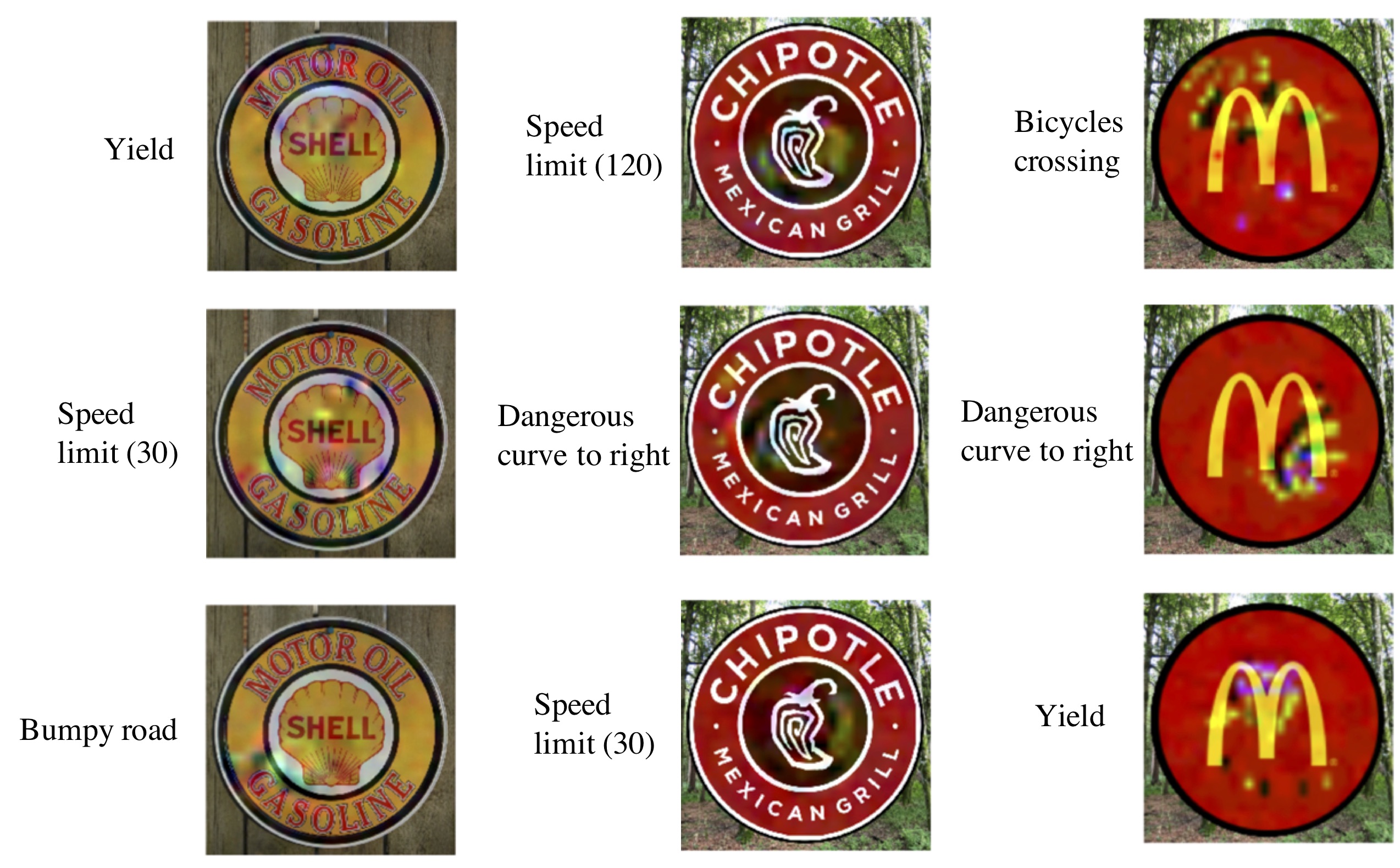}\label{fig: logo_adv}} \\
	\subfloat[Some samples of our Out-of-Distribution (Custom Sign) attack, along with the labels they are classified as.]{\includegraphics[width=0.54\linewidth]{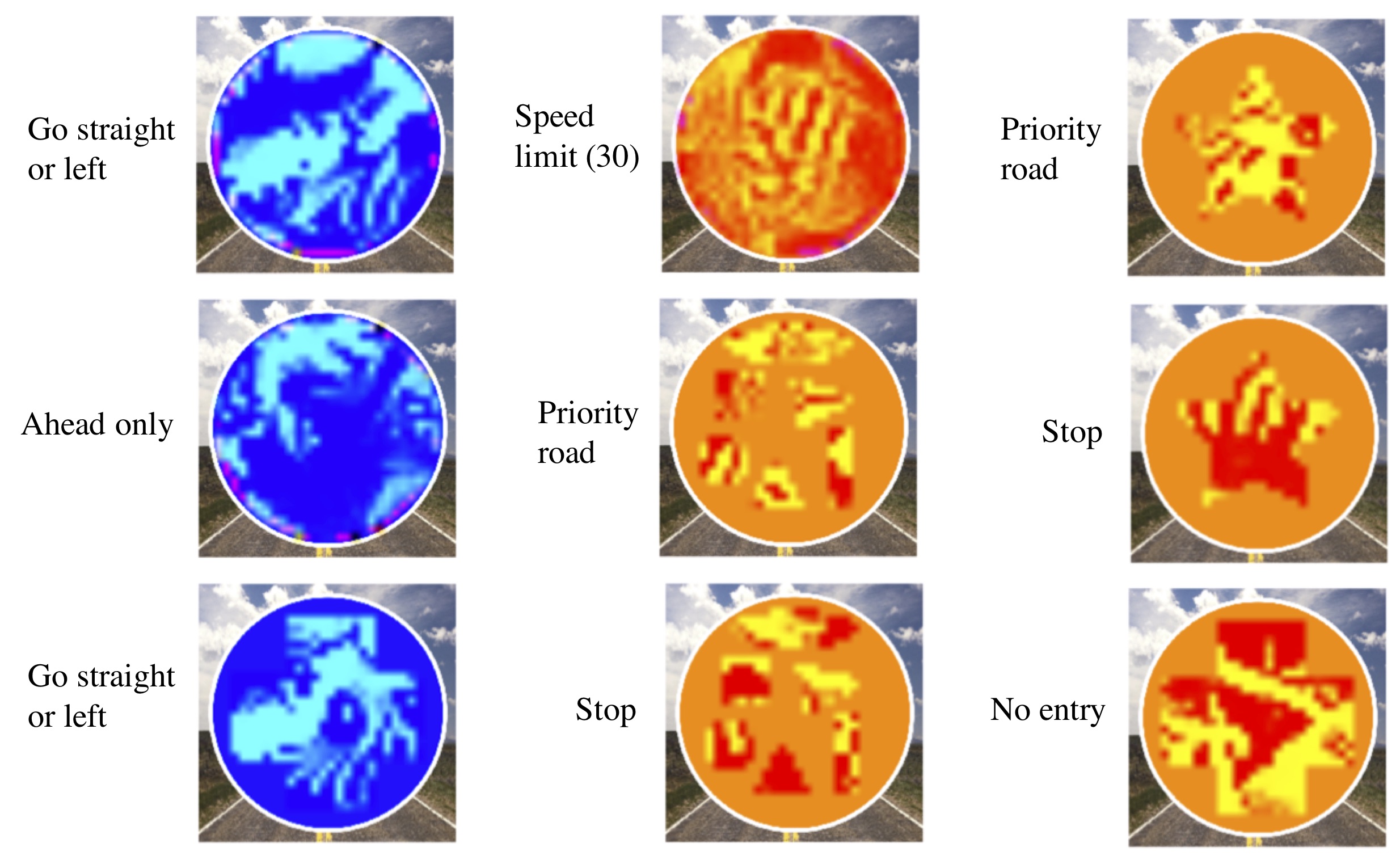}\label{fig: custom_adv}}
	\caption{Adversarial examples that achieve SPAS of 100\%.}
	\label{fig: extra_examples}
\end{figure*}

Here, we include some of the adversarial examples we have generated for all of the three attacks in Figure \ref{fig: extra_examples}. The samples shown also achieve 100\% SPAS.

\section{Appendix: Adversarial training} \label{app: adv_training}
Ideally, the $\xad$ in Eq.\eqref{eq: adv_training} (modified loss function) should also be generated in the same manner as the robust physical attacks. However, as mentioned earlier, each adversarial example generated using our method takes around 60s to generate. This makes it impractical to run the optimization for every training batch.

The simplest attack against DNNs is known as the Fast Gradient Sign (FGS) \cite{goodfellow2014explaining} attack which is an \emph{untargeted attack} that involves adding a perturbation proportional to the sign of the gradient of the loss function of the neural network $ \xad = \bfx + \epsilon \text{sign}(\nabla_{\bfx}\ell_f(\bfx,y)).$
The advantage of this attack is that it is extremely fast to carry out so it can be incorporated into the training phase of neural networks to make them more robust to adversarial examples. 

\noindent \textbf{Note about norms:} While our attack samples have on average a maximum $L_1$ perturbation of around 30, the $L_1$ ball with that radius is too large to train with, since the ball has to lie in the unit hypercube. Since most of the adversarial examples do not modify each individual pixel by more than 0.3, it is a reasonable upper limit to use while training. In our experiments, we tried training with adversarial examples constrained with the $L_1$, $L_2$ and $L_{\infty}$ norms and found the defense to work best with an $L_{\infty}$ norm constraint.

We also tried using the iterative adversarial training defense proposed by Madry et al. \cite{madry_towards_2017}. However, we faced a number of issues with this defense. First, it is known that the adversarial loss with iterative adversarial samples does not converge for classifiers with low capacity. When we attempted to train the \mltscl with iterative adversarial examples and the augmented training data, we observed the same behavior. Using the standard training data, the model converged to a validation accuracy of 96.2\%. However, its performance on adversarial examples generated using the \auxdata was inferior to that of the \mltscl with standard adversarial training.

\end{document}